# The COVID-19 Pandemic on the Turkish Twittersphere

Burak Ozturan

June 28, 2022

**Abstract**

With the increase in the time spent at home, social media platforms' role have become an integral part of the public discussion in the COVID-19 period. Individuals use social media platforms to express their emotions, interact with one another, and engage in public debate. Therefore, it is essential to analyze social media platforms for those who desire to understand public opinion during the pandemic. This thesis is the first study that examines the Turkish Twitter sphere to understand the change in public opinion during the COVID-19 outbreak. For that purpose, starting from the 12th of February, 2020 (one month before the first announced coronavirus cases in Turkey), 4.3 million Turkish tweets with a broad range of keywords are collected until June 2020 to investigate the public opinion change in different topics and to examine the actors leading to that change. The scope of the analysis is not only health-related discussion but also includes a broader range of themes such as politics, economy, and disinformation. This study also collects 4.15 million Turkish tweets with keywords of vaccine ("aşı" in Turkish) from 4 April 2020 until 17 March 2021 to unpack the health of the information ecosystem. Preliminary results suggest that (i) religion is the prominent phenomenon for Turkish people's perception of the pandemic, (ii) and the Turkish Twitter-sphere is highly vulnerable to mis/disinformation operations and (iii) several communities with divergent interest exist in the vaccine network.

# Contents

# List of Tables

# List of Figures

# 1 Introduction

COVID-19 pandemic has been influencing the world since the first months of 2020. It started as an epidemic but soon turned into a pandemic with its high rate of spread. Countries have taken serious measures such as lock-downs and curfews to reduce the virus spread rate while encountering drastic societal, economic and political problems. Those restrictions have led people to spend their time at home. Willingly or not, individuals have started to spend much more time at home and use social media platforms frequently to voice their current experiences, articulate their opinions, evaluate the government health policies and get socialized[1]. Therefore, social media platforms provide abundant data for analyzing human behaviour in the pandemic. By understanding the content of social media posts, researchers might deduce the behavioural response of individuals to a particular stimulus, governments can optimize their policies in the future.

There have been many recent studies about COVID-19 pandemic by using Twitter data. [1, 2, 3, 4, 5, 6, 7]. These studies use coronavirus-related keywords such as COVID-19, COVID, coronavirus, mask, and lockdown to create the Twitter dataset. While collecting hundreds of millions of tweets, these datasets exclude those tweets that do not contain any of the coronavirus-related keywords. However, the excluded tweets were also created during the pandemic so that they contain contextual information about the public opinion during the pandemic. In this regard, more comprehensive keyword list is needed to understand the changes in public opinion better. Moreover, although some keywords that address coronavirus such as COVID-19, COVID are used ubiquitously, tweets may have some language-specific spellings and words rather than uni-versal terminology. For example, Turkish Twitter are using some alternatives on Twitter, such as "corona" and "coronavirus", which are as often used as the universal English COVID-19 variations. These reasons necessitate the curation of a more comprehensive dataset to investigate the dynamics of the Turkish public in the pandemic.

[1] https://www.statista.com/topics/7863/social-media-use-during-coronavirus-covid-19-worldwide/

This study fills this gap by leveraging a broad range of language-specific keywords list identified for six important public debate themes most likely to be affected during the pandemic. Those themes are health, politics, international relations, economics, religion, and disinformation. Starting from one month before the first COVID case in Turkey, this study has collected 4.3 million Turkish tweets to analyze the change in the Turkish Twitter users' public opinion of six essential public debate themes and the dynamics of the Twittersphere during that time.

Concurrently, social media platforms can be subjected to increased misinformation and disinformation activities, especially in times of crisis. [8]. This phenomenon is now described as infodemic [9], meaning that false contents surround the information ecosystem, and it is increasingly troublesome for individuals to find reliable information. Infodemic in the crisis times such as COVID-19 pandemic and natural disasters is particularly problematic since it can cost people's lives. Therefore, it is essential to understand the characteristics of these malicious activities and major actors' behaviour to take measures to reduce the spread of false content and the belief in that content. By analyzing the content and the network of the 4.15 million vaccine tweets, this study explores the actors and communities and their agendas during the pandemic.

## 1.1 Research Questions & Contributions of This Work

This study investigates three main research questions by leveraging an extensive COVID-19 Turkish Twitter dataset.

- How did the public debate on politics, economy, health, religion and disinformation change during the pandemic? What were the topics that dominated the discussion before the COVID-19 pandemic? What have been the new emergent topics subsequently?

- What kind of rationale do the Turkish people use to perceive the epidemic? Do they discern the problem as a health problem, or do they have other tendencies to explain the pandemic outbreak? This question is important to identify patterns of potential mis/disinformation patterns that infiltrate social media platforms.

- What types of actors and networks do exist in the Turkish Twittersphere during the pandemic? Who are the most active accounts in that period? How do the agendas of these users differ from each other?

This study provides the following important contributions to the social media studies and Turkish Twitter studies in particular:

- It offers a new research design to investigate changes in public opinion during major events such as the pandemic. It tracks the public opinion change on six essential themes: politics, international politics, health, economy, religion, disinformation rather than only coronavirus keywords [1, 2, 3, 4, 5, 6, 7]. In this way, we can obtain broader insights into public opinion change.

  By the detailed keyword list, it analyzes the public opinion change on six essential themes: politics, international politics, health, economy, religion, disinformation.

- This study also unpacks the Turkish Twitter information ecosystem during the COVID-19 pandemic with the analysis of the content of the tweets, analysis of the links and the network of the Twitter users. Given the fact that Turkey is the 6th country based on the number of Twitter users as of July 2020[2], understanding the dynamics of Turkish Twitter can provide important insights for comparative Twitter studies.

- Importantly, this study provides 8.5 million Turkish tweets as a public dataset. Researchers can use the dataset to answer important social scientific questions. For instance, researchers can investigate the change in trust and public support towards leaders, politicians, or health ministers by combining the named entity recognition (NER) and the sentiment analysis algorithms.

The following section describes the related literature of social media data during the pandemic and mis/disinformation. Section 3 presents the research design of the study Turkish Twittersphere public opinion. It, first, explains the creation of the keyword lists for each theme. Then, it articulates the results of the public opinion change. Section 4 illustrates the vaccine information ecosystem on the Turkish Twittersphere. Initially, it analyzes the content of the tweets by implementing quantitative text analysis techniques. Subsequently, it defines the communities of vaccine tweets network and explores the agendas of the different communities. Lastly, section 5, discuss the finding and explicates the future direction of this study.

[2]https://www.statista.com/statistics/242606/number-of-active-twitter-users-in-selected-countries/

## 2 Related Literature

There has already been a considerable amount of datasets that collect COVID-19 related data from social media. For instance, Banda, Tekumalla, Wang, Yu, Liu, Ding, and Aremova curated dataset of over 152 million tweets [10]. Chen, Lerman, and Ferrara have collected 123 million tweets (60% of English) and have a closer look at the characteristics of Japanese, Italian, and Spanish Twittersphere [1]. Qazi, Imran, and Offli curated 524 million geo-tagged multilingual tweets with keywords basics COVID-19 keywords variations like COVID-19, mask and social distancing with country names [11].

These studies use mostly the COVID-19 keyword variations to flow the tweets into their dataset. Although they could catch the multilingual tweets with universal COVID-19 keywords, they systematically exclude the other tweets that have been created during the pandemic but do not incorporate any COVID-19 keywords. Therefore, the scope of these keyword lists should be expanded to analyze the public opinion in-depth and tailored for comprehensive country-specific research.

The mis/disinformation literature has increased after the 2016 United States presidential election. Bessi and Ferrara find that one-fifth of Twitter conversation is generated by social media bots during the presidential election [12]. Those accounts could disrupt the public discussion and challenge the confidence in democracy [12]. And, Ferrara analyzes 43.3M English tweets about COVID-19 and provides evidence of the use of bots presence in the COVID discussion. They promote political conspiracies in the United States [13].

To investigate the spread of false content online, Vosoughi, Roy, and Aral analyze 126,000 stories shared 4.5 million times by approximately 2 million users. They find that false content is shared more rapidly than the credible ones [14]. Grinberg, Joseph, Friedland, Thompson, and Lazer show that interacting with fake news sources on Twitter is not uniform in the Twitter population but concentrated in a tiny segment of Twitter users [15]. And Shao, Ciampaglia, Varol, Yang, Flammini, and Menczer find that those contents from low-credibility sources are dominantly amplified by social bots [16].

There is also burgeoning literature regarding health misinformation. Liang, Fung, Tse, Yin, Chan, Pechta, Smith, Lameda, Meltzer, Lubell, and Fu implement social network analysis to ebola messages on Twitter to investigate the spread of Ebola messages and identify the influential users in that network [17]. Singh, Bode, Budak, Kawintiranon, Padden, Vraga investigates the URLs shared on Twitter in the discussion of COVID-19. They make a distinction between reliable health sources, traditional news sources and low profile misinformation sources. They also find that low quality is shared more than reliable sources [18]. Pulido, Carballido, Sama, and Gomez analyze the spread of fake news and credible contents during the COVID-19. They find that false content is tweeted more but retweeted less than evidence-based content. They also show that people interact more with fact-checking than the fact itself [3]. On the other hand, Cinelli investigates the diffusion of COVID-19 related information on social media platforms, Twitter, Instagram, YouTube, Reddit and Gab. They analyze the engagement in the COVID-19 content and evaluate the discourse for each platform. They find that information spread patterns of reliable and questionable sources are not significantly different [7].

To show the impact of low-quality information on public health, Gallotti,

Valle, Castaldo, Sacco, and Domenico analyze 100 million tweets in 64 languages during the COVID-19 pandemic. They find that low-quality information threats public health via causing the irrational social behaviour [6]. Thompson and Lazer also review the current information ecosystem discussing how personal access to health information with the internet influence the public health outcome. They also propose important recommendations to improve the circumstances of the online information ecosystem [19].

# 3 COVID-19 Turkish Twittersphere Public Opinion

With the increase in the time spent home, individuals use social media platforms to engage in public discussion. For instance, they express their opinion about the pandemic restrictions, their current psychology and the problems with the economic turbulence. Therefore, it has vital importance to study social media platforms to understand the public opinion and public opinion change caused by the COVID-19. Turkey, with 61% of the Twitter penetration rate, is among the countries where the Twitter penetration is significant.[3]

This section analyzes the public discussion that took place on the Turkish Twittersphere during COVID-19. First, the methodology of the study will be presented. Subsequently, I will analyze the results of the study.

## 3.1 Methodology

To analyze the public opinion on Turkish Twittersphere during the pandemic, I determine six important socioeconomic and political themes that are frequently discussed on Turkish Twitterpshere: COVID-19, (domestic) politics, religion, economy, international relations, and disinformation.

[3] https://www.statista.com/statistics/284503/turkey-social-network-penetration/

The time interval of the dataset is from the 12th of February 2020 until 08 June 2020. However, this study illustrates the public opinion change between one month before the first official COVID case in Turkey ( 11 March 2020 ) and the month later after the outbreak ( 11 April 2020 ). The time interval for this study is set for two months to reduce the complexities in managing big data since there has been high volume of social media posting during those period.

Only Tweet IDs can be stored publicly due to Twitter's Terms of Service. While extracting the ID's, repetitive tweets were cleaned. As a result, the dataset consists of unique tweets for each of 6 theme. The IDs of the tweets are publicly stored in the GitHub page of the study.[4] as per the Twitter's Terms of Service.

For the replication or another research using those tweets, tools for hydration/dehydration such as Hydrator[5] and Twarc[6] might be convenient tools to obtain the tweets of this study. But it should be noted that the deleted contents will not be available.

In following subsections, I elaborate on the data collection and analysis techniques and later on I present results of the study.

### 3.1.1 Extracting keywords and dividing them into themes

I collect a comprehensive list of keywords for the six themes: COVID-19, domestic politics, international relations, disinformation, religion, and economy

[4] https://github.com/burakozturan/css_covid19/tree/master/id
[5] https://github.com/DocNow/hydrator
[6] https://github.com/DocNow/twarc

with the intense Twitter search. The English translation of these keywords are as follows:

- *COVID-19 Turkish Variations*:

  COVID-19, Covid-19, Covid 19, Kovid-19, korona, koronavirus, virus, corona

- *Politics:* President, Ministry of Health, Fahreddin Koca (Minister of Health), War, Minister, Campaign, Peace Spring (Turkish Army mobilization into Syria)

- *Foreign Countries/International Relations:* China, Europe, US, England, Italy, Spain, Germany, France, Japan, South Korea, Iran, Israel, WHO, NATO

- *Disinformation:* Unfounded, Fake, Provocative, Bragging, Conspiracy, Game, Big Game, Zionism

- *Religion:* Ministry of Religious Affairs , Mosque, Friday, Amulet, Religion, Adhan, Sela, Patience , Calamity

- *Economy:* Credit, Scholarship, Unemployed, Unemployment, Cash, Support, Solidarity, Food, Struggle, Officer, Company, Customs

Correspondingly, the tweets are collected under the each six themes separately, so that researchers do not have to divide the data according to themes as subjects for their work. They can focus on the data for one particular theme or any combination of themes they desire.

Keywords are selected with a random Twitter search for each theme. Instead of tracking one particular keyword, I aim to incorporate the relevant keywords

as much as possible for the study period to comprehensively capture the change in public opinion. Although those keywords list is not exclusively complete for the particular theme, and some of the keywords would be time-variant, it still offers a novel research design to analyze the public opinion change during a crisis.

### 3.1.2 Codes for Analysis

Python is used for data collection, and analysis. After collecting the tweets with each keyword, I created a thematic dataset by combining the tweets under the same theme. For instance, the tweets with keywords religion and mosque belong to the same thematic dataset religion.

Tweets are pre-processed by using Natural Language Tool Kit (NLTK) package and manual coding for Turkish language adaptation. For example, I create a Turkish stopwords list (words does not convey important contextual information such as the, one, this) specific to COVID-19 context that does not exist before.

For the visualization of the public opinion change, statistical frequency analysis of the words and word cloud techniques are used. You can find the Jupyter Notebooks with detailed explanations here.[7]

## 3.2 Results

This section, firstly, presents the descriptive statistics of the tweets. Secondly, the most frequent ten words before and after the pandemic outbreak are illustrated to show the contextual continuity and/or breaks in six themes. A more detailed analysis will be the subject of the future research.

[7] https://github.com/burakozturan/css_covid19/tree/master/codes

### 3.2.1 Descriptive Statistics

Table 1 shows the tweets for all the themes for two months, between 12 February and 11 March 2020. Before period is from 12 February till the first official COVID-19 case in Turkey, 10 March 2020. After period covers the tweets from the 11 March until 8 June 2020. Figure 1 shows the daily COVID-19 theme tweets with key events that affect the number of tweets.

For all the themes, the number of tweets created in the after period is higher than the number of tweets in the before period. In total, approximately 4.3 million tweets are collected separately for every six themes as can be seen in Table 1. The tweets within each theme are unique. However, there are some overlaps across tweets between themes. Once the tweets across themes merged, 3.5 million unique tweets exist in the dataset.

| Theme | Before | After | Total |
|---|---|---|---|
| COVID-19 | 161,109 | 1,427,569 | 1,588,678 |
| Politics | 355,992 | 367,992 | 723,984 |
| Disinformation | 51,521 | 88,007 | 139,528 |
| Religion | 215,699 | 436,309 | 652,008 |
| Economy | 19,532 | 74,682 | 94,214 |
| International Relations | 326,030 | 845,428 | 1,171,458 |
| **Total** | 1,129,883 | 3,239,987 | **4,369,870** |

Table 1: Number of Tweets for Themes between 12 February - 11 April

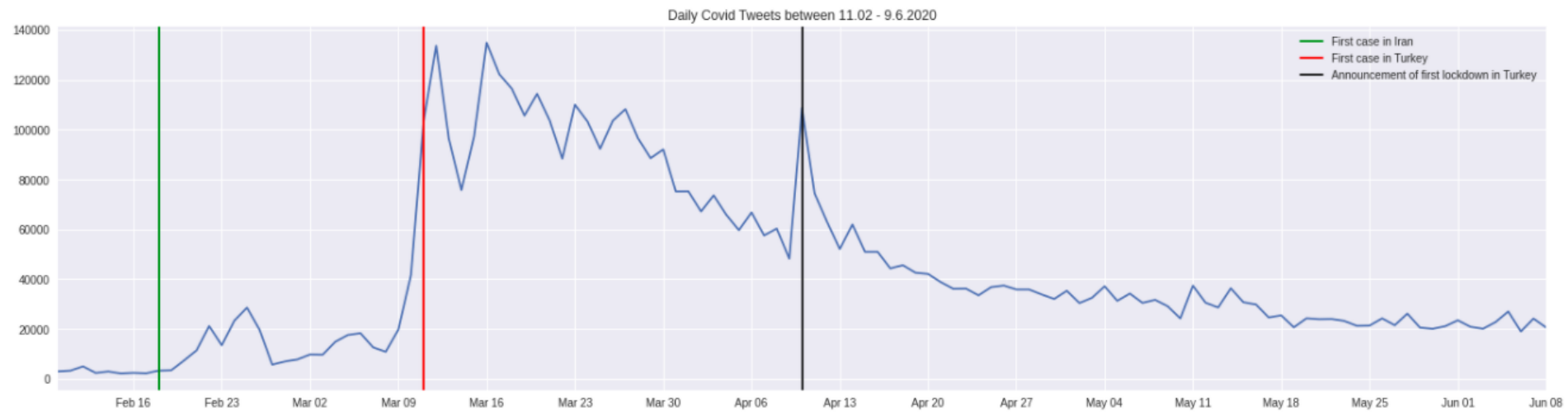

Figure 1: Daily COVID-19 tweets between 12 February - 08 June

### 3.2.2 Main Findings

This section will first elaborate on the change in the daily COVID-19 tweets between 12 February - 08 June. Then, I investigate the public opinion change for the six themes COVID-19, domestic politics, religion, economy, international relations and disinformation in detail by concentrating on the before and after one month window (12 February - 11 April).

Figure 1 illustrates daily tweets from 12 February onward, a week before the first Iranian case. After the Iranian case, shown by the green vertical line, Turkish tweets about COVID-19 increased and reached their peak one day after the first Turkish case, shown by the red line. Later on, it continued downwards with an increase on the announcement of the first lockdown in Turkey.

Figure 2 indicates frequent references were made to countries such as Iran, Italy (italya) and China (Çin), where the COVID-19 cases were the most severe before the pandemic outbreak. Afterwards, the words "Turkey" increased dramatically, and the words "God" (Allah) and "health" (sağlık) come to the fore. This suggests that the focus of Turkish Twitter users have shifted towards the pandemic at home. Importantly, the prevalence of the word "God" give insights that Turkish people have been using religious arguments to perceive the pandemic. Qualitative analysis of those tweets with the word "God" shows that people regard the COVID-19 pandemic as a divine punishment and expect salvation from "God".

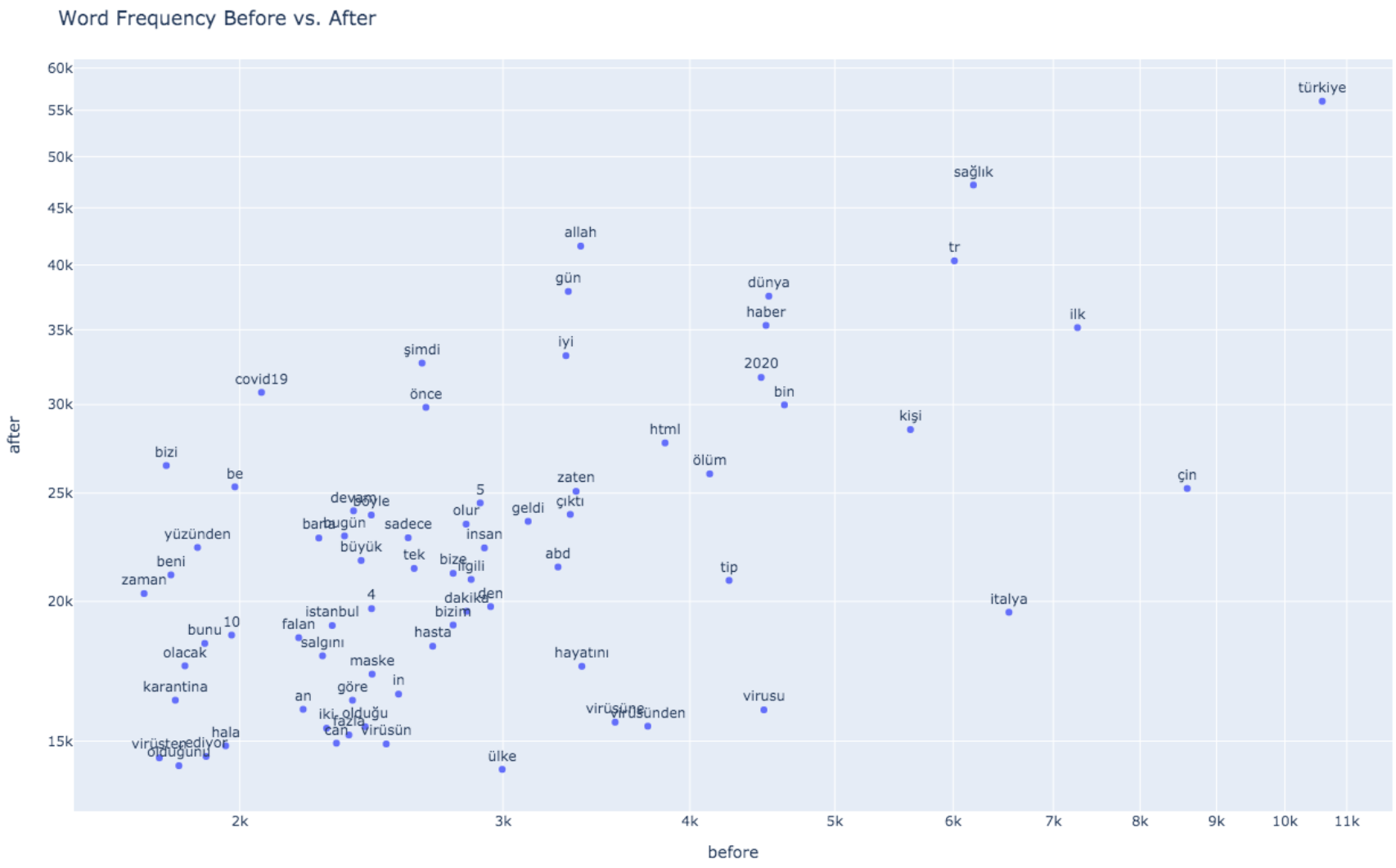

Figure 2: Most Frequent 10 words for COVID-19 Before & After

Figure 3 shows the public opinion changes in the (domestic) politics theme. Before, the focus was on the refugee crisis (mülteci) that the Syrian refugees in Turkey were transported to the Greece border and Turkey's military operation in Idlib (asker, savaş). Then there has been a significant shift from those topics to science and health (drfahrettinkoca, the twitter handle of health minister), as can be seen at the right graph in Figure 3.

Figure 4 presents the change in religion theme. The emphasis of the religious discussion before was on Turkish soldiers in Idlib, Syria. Twitter users wish to rest (rahmet) for martyrs' souls from God and patience (sabır) to their families. Later on, the focus has concentrated on the cancellation of prayer (cuma) in mosques as a pandemic precaution. Also, the relative increase in the frequency of the word patience (sabır) gives some insights that people expect the pandemic to disappear if they somehow bear with it for some time.

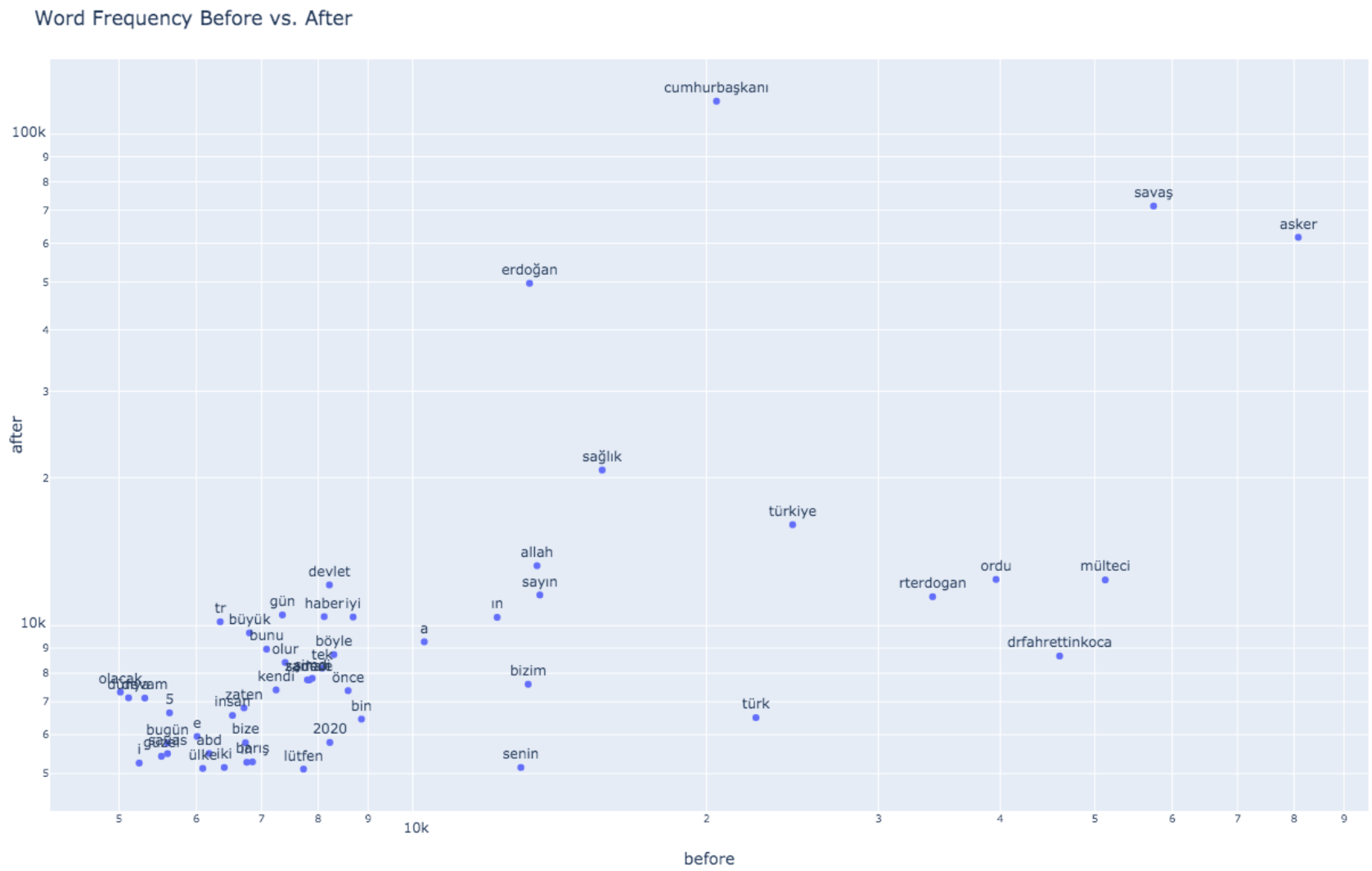

Figure 3: Most Frequent 10 words for Politics Before & After

Figure 5 illustrates the public discussion about the economy. Previously, the dollar's value against Turkish lira and unemployment were distinguished. Subsequently, the words "lockdown" and "state" (devlet) emerged prominently. The prominent use of the word "state" after the pandemic is an important indicator for Turkish Twitter users' expectancy to see the state as a key actor to solve the economic difficulties pandemic causes.

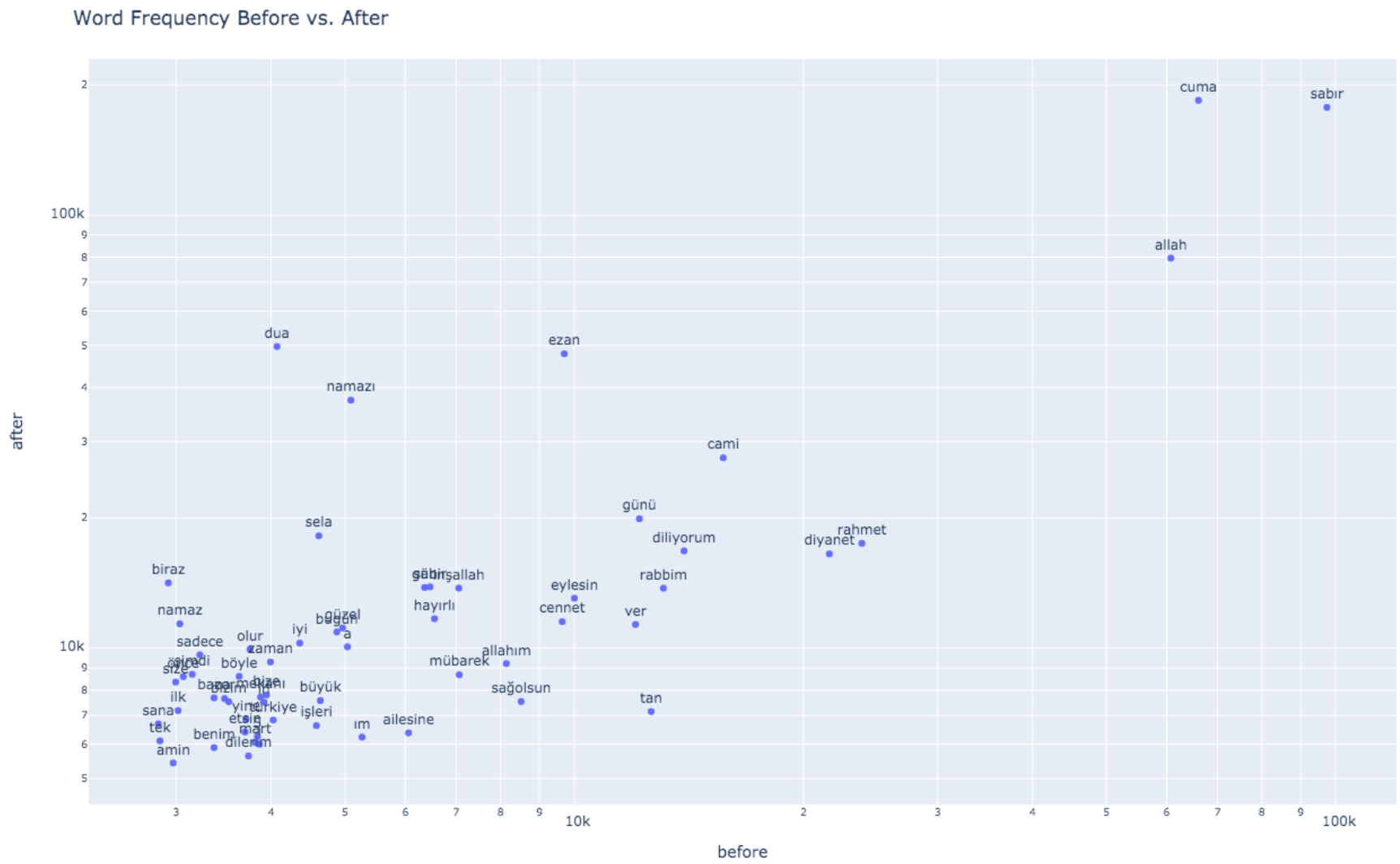

Figure 4: Most Frequent 10 words for Religion Before & After

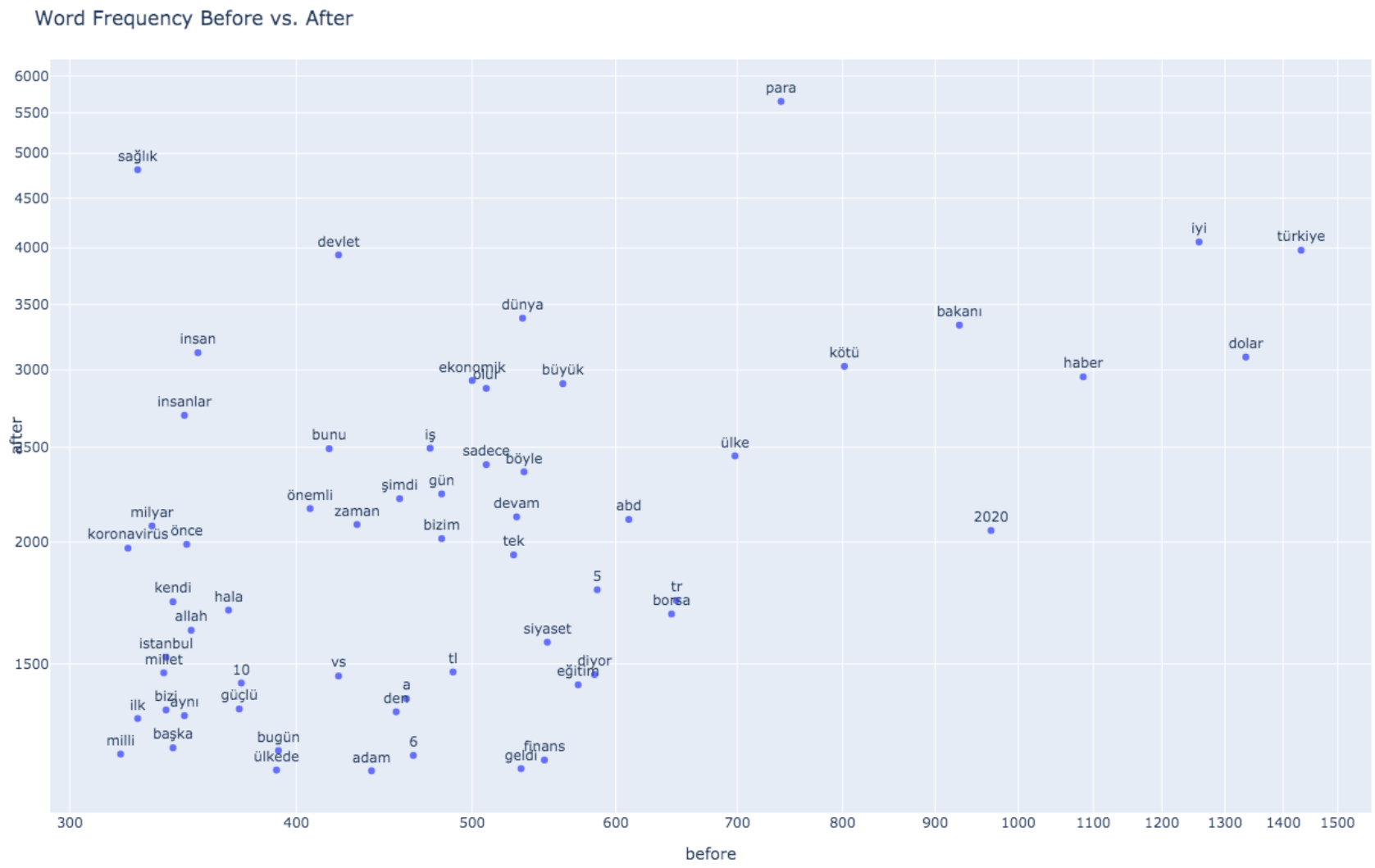

Figure 5: Most Frequent 10 words for Economy Before & After

Figure 6 demonstrates changes in the theme of international relations. Before the pandemic started, Turkey's military operation in Syria, where they encountered Russian forces, was referenced. After, the COVID instances in European countries mostly occupy the conversation.

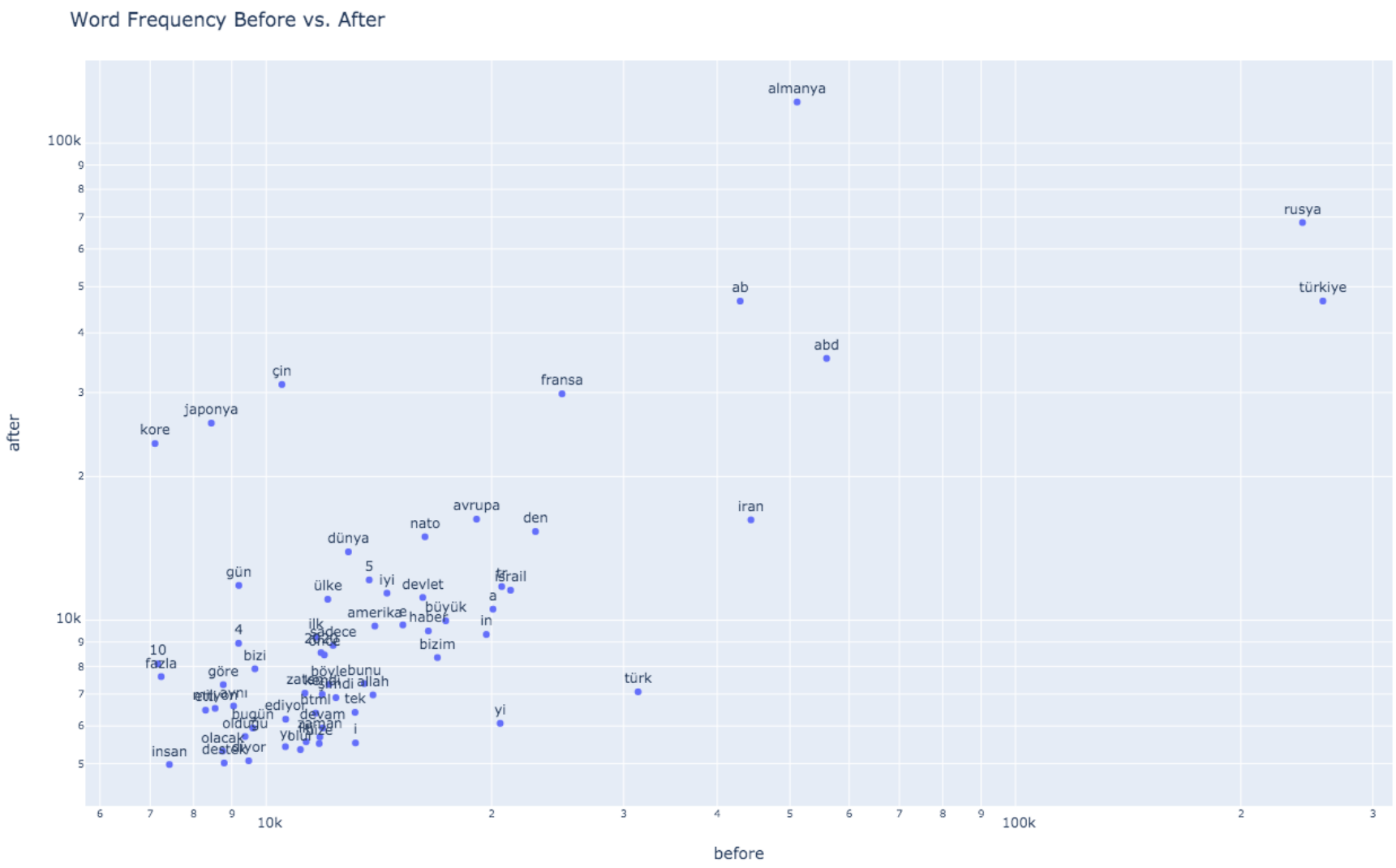

Figure 6: Most Frequent 10 words for International Relations Before & After

Figure 7 shows that the word "fake" has been always present in both before and after the outbreak. Furthermore, there has been a significant increase in the word "conspiracy". These two observations suggest that Turkish Twittersphere is vulnerable to mis/disinformation activities. The word cloud of the themes can be found at Appendix C.

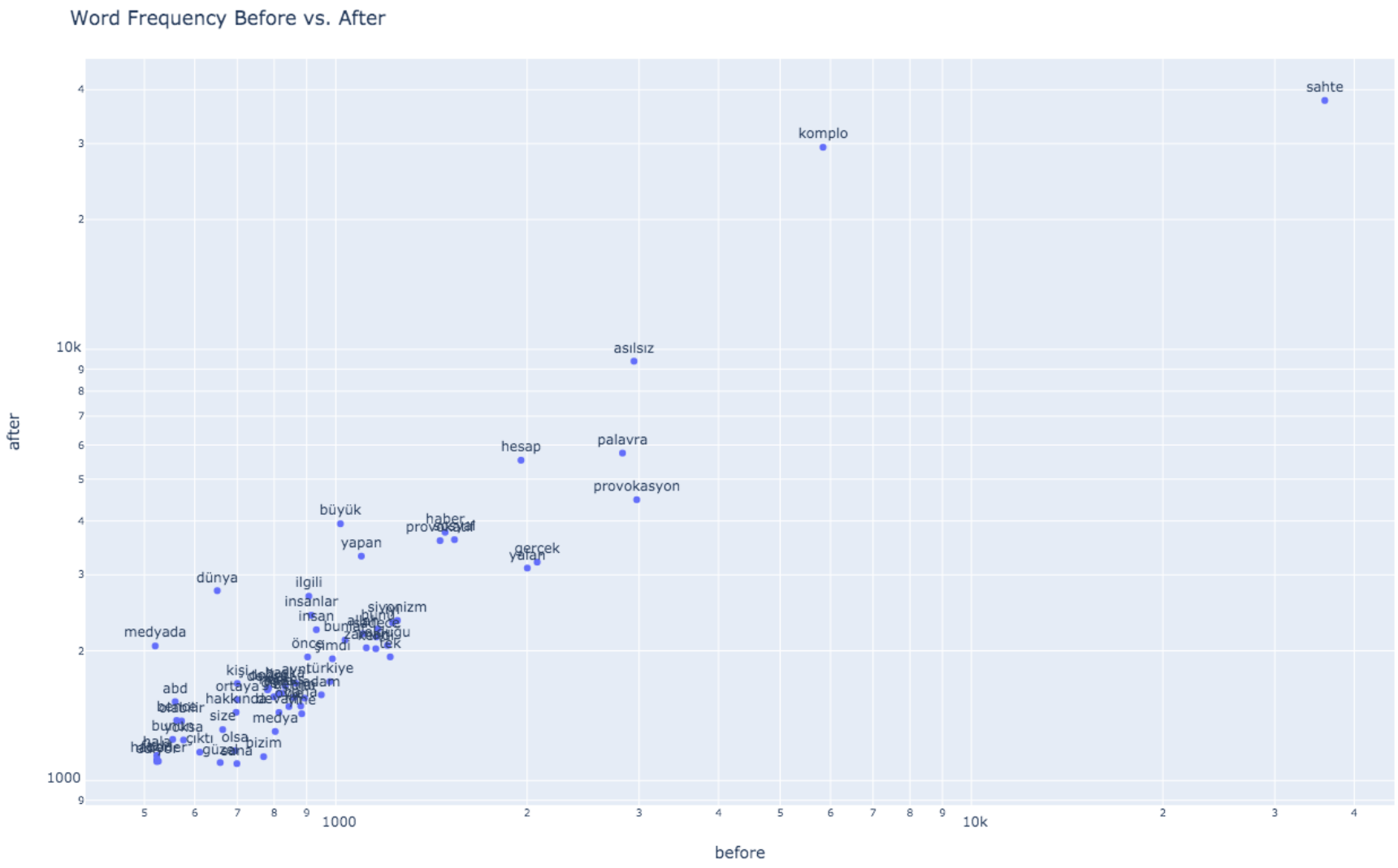

Figure 7: Most Frequent 10 words for Disinformation Before & After

### 3.2.3 Twitter Information Ecosystem

This study uses the Botometer score to identify the social bots. Botometer is a machine learning system designed to detect social bots by using the linguistics, temporal, and network features of twitter accounts. It gives the likelihood of a particular account that can be classified as bots. Specifically, If the botometer score is higher than 0.5, then the account is classified as bot. [20]. Result of this study that was obtained in 2020 used Botometer v3 [21].

Figure 8 shows the Botometer score distribution of the most active 15,000 accounts between 12 February and 8 June. These 15,000 accounts are 1% of the total active accounts during that period. They created 25% of 5 million tweets. Accordingly, we can conclude that the Turkish Twittersphere is highly concentrated during the pandemic regarding the number of tweet creators. As Ferrara [13] suggests, for the US Twittersphere, Turkish Twitter users might also be classified into 3 groups: social bots, news agencies, and real humans.

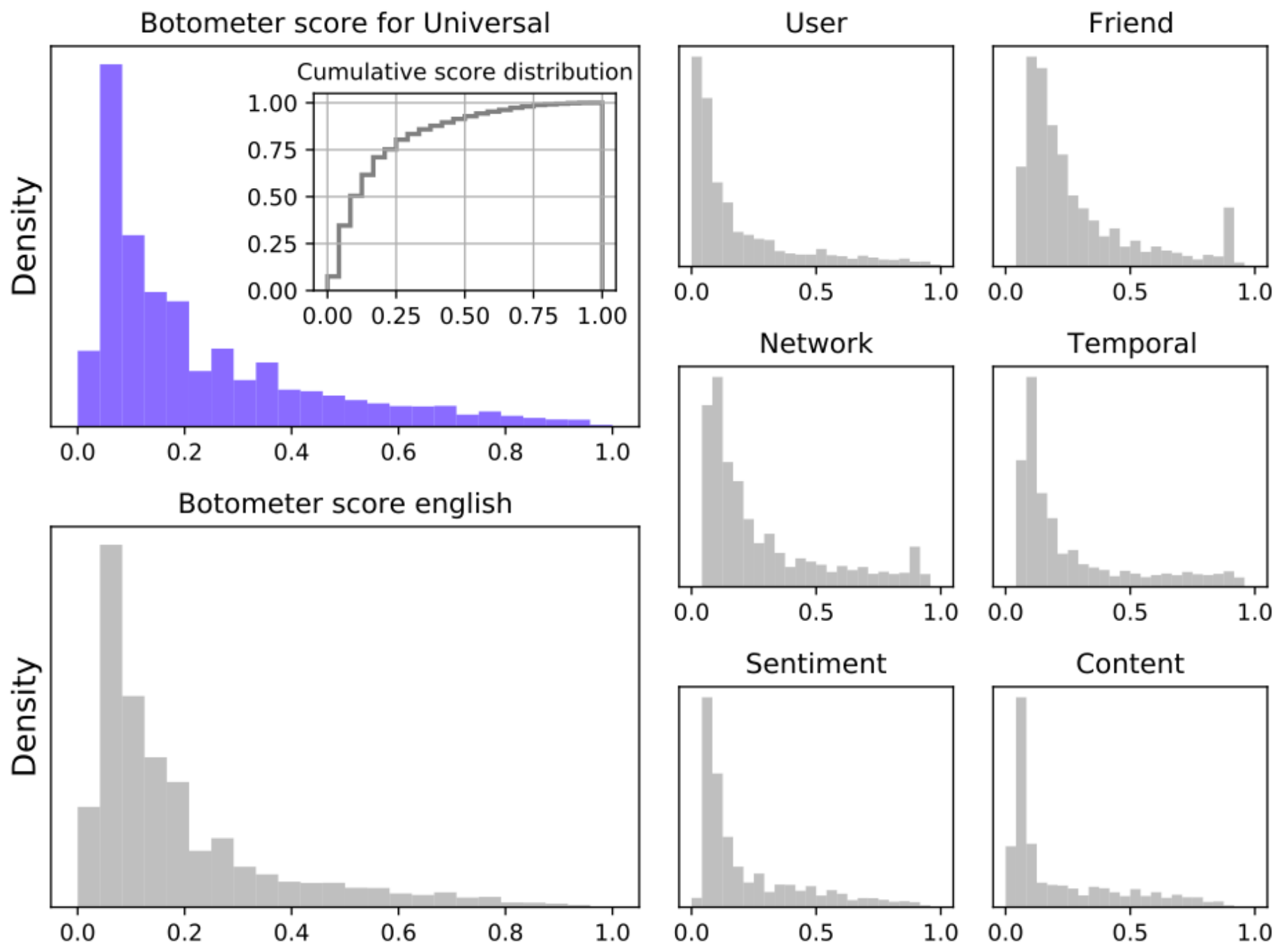

Figure 8: Botometer Score Distribution of the most active 15,000 accounts

The top left of Figure 8 shows the cumulative distribution of the bots score. Approximately 10% of the most active 15,0000 accounts are social bots. This result is also compatible with the US case [22]. The other subfigures with a grey bar indicate the botometer score density for utilizing only one feature: sentiment, content, and account network.

It is crucial to analyze the characteristics of the bots whether they have been able to distort COVID discussion on Turkish Twittersphere. By investigating this issue and spread patterns of the online content during health emergencies would provide important insight to combat future disinformation campaigns leading to infodemics [9].

# 4 Turkish Vaccine Information Ecosystem on Twitter

## 4.1 Daily Tweets

There are 4.15 million tweets during the study period from 4 April 2020 until 28 March 2021. Figure 9 shows the logarithm of the number of daily Turkish tweets containing the keyword vaccine ("aşı" in Turkish), biontech and sinovac. Initially, the volume of tweets that contain the vaccine (aşı) keyword was quite low, and it changes depending on the news about the vaccines that report the phase studies of different kinds of vaccines. However, there were three significant increases in total volume. The first increment started on 3 December 2020. Minister of Health (Dr Fahrettin Koca) declared the vaccination schedule: when and to whom the first vaccination will start on that day. With this important development, daily Twitter activity regarding the vaccination reached 60,000 levels. After 2 weeks, the Turkish Twittersphere made another peak on 17 December. On that day, vaccination started in Europe, and Turkish Twitter users put this event on their agenda, and daily Twitter activity increased to 70,000 levels. The third and the highest ramping up started on o the first COVID-19 vaccination in Turkey on 14 January 2021. They reached their peak on 15 January 2021 - when some of the political party leaders were vaccinated- by more than 162,000.

On the other hand, intuitively, the volume of tweets with biontech and sinovac was lower than the tweets with vaccine keywords in general, even though there was one exception. The biontech tweets outperformed the others once the company announced the vaccine's phase 3 study results.

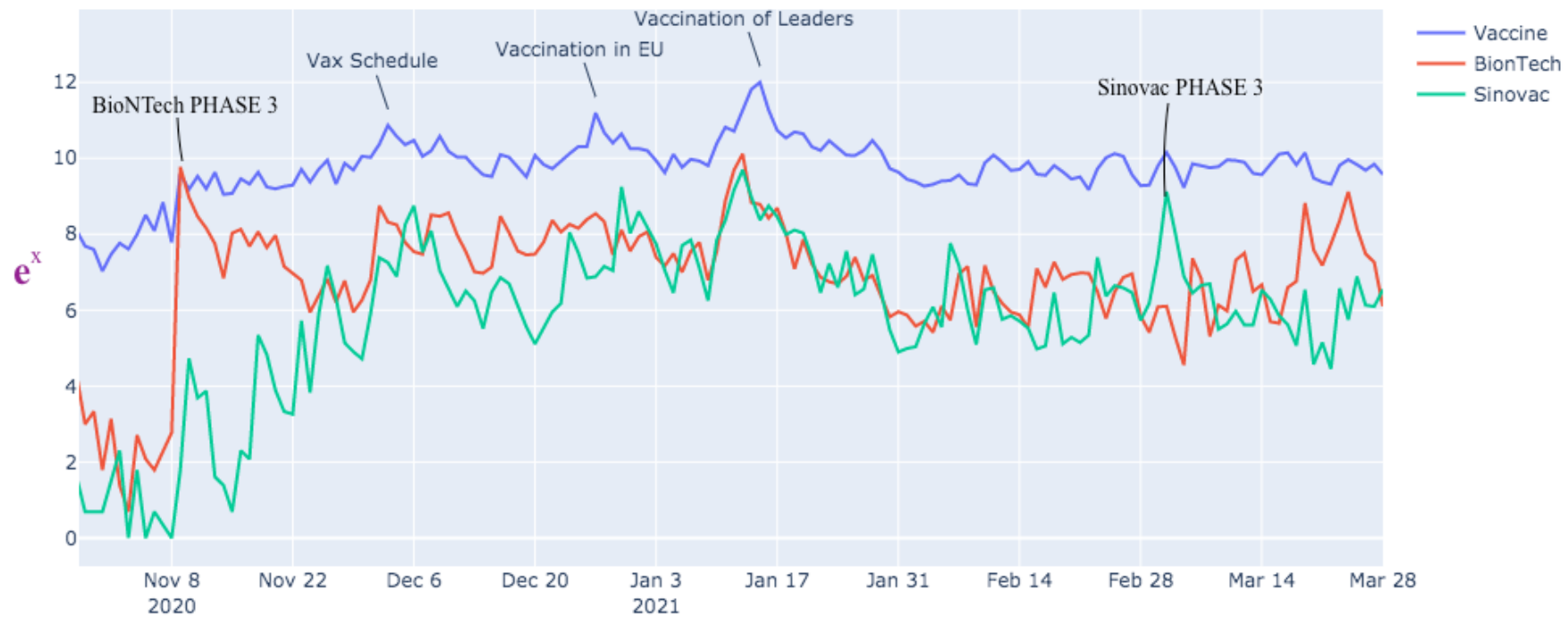

Figure 9: Daily Natural Logarithm of Number of Tweets

## 4.2 Methodology

The full-archive search endpoint of The Twitter API product track for academic research allows researchers to collect historical tweets, dating back to March 2006 (Twitter's establishment date). Importantly, to collect a complete tweets archive rather than a sample of tweets. Using academic Twitter API full-archive search, I have collected all the tweets from 4 April 2020 until 28 March 2021 with the keywords "aşı" ("vaccine" in Turkish), "biontech" (BioNTech, German biotechnological company that produce the COVID-19 vaccine with Pfizer) and "sinovac" (Sinovac, Chinese bio-pharmaceutical vaccine producer company). Twitter's search function can automatically capture the lowercase, uppercase and English keyboard adaptation of the Turkish specific characters. Tweets are preprocessed by common natural language processing pipelines. I first convert all the tweets to lowercase and clean stopwords and punctuations. I utilize unigram and bigram tokenization together for the frequency analysis of the tweets.

In order to analyze the links that were shared on the Turkish Twittersphere during the period of study, first, I filter the tweets which incorporate URL entities. Subsequently, I retrieved unwound versions of the links that are not implemented any shortening and indicates the full link. Then, I extract the top domains shared during the study period by combining the individual links with the same domain addresses.

I used berturk-social-5m [23] for sentiment analysis of Turkish tweets with the keywords of "sinovac" and "biontech". The model uses RoBERTa[24] language model that implements some extensions to perform faster than the traditional BERT model[25]. The model is trained with five million Turkish tweets and fine-tuned for the sentiment analysis with six thousand tweets. As an output, it classifies tweets as either positive (POS) or negative (NEG) or neutral (NOTR). I convert these outputs to +1, 0 and -1, respectively, to extract the average sentiment of the tweets. The model does not have any accuracy benchmark accuracy for the sentiment analysis task. To evaluate the model's performance, I perform hand coding for the sentiments of ten tweets for each of the model's output classes. Even though the model has effective performance on classifying negative and neutral sentiments, it does not have the same adequate performance lever for the positive tweets (see Appendix A).

To analyze the actors and relations between the different communities in Turkish Twittersphere, I created three directed network maps of the tweets containing the keywords vaccine, biontech and sinovac separately. For this purpose, each user actively retweeted, mentioned, quoted someone, and/or passively interacting with someone with at least one kind of interaction serves as a network node. For edge-wise, I combine all the interaction types between users on Twit-ter (retweet, mention, quotation) into one by summing. For example, if node A retweeted node B twice and quoted once, the edge weight between A and B is 3. Similarly, if node B quoted node C three times and mentioned node C only once, the edge weight between B and C is 4. To reduce computational complexity, I exclude the edges whose weight is less than 3. Subsequently, I use the ForceAtlas2 algorithm [26] for visualization layout. Also, I use modularity [27] to detect communities on the network map and eigenvector centrality to identify the influential nodes in the network.

## 4.3 Main Findings

This section describes the main findings of the study. First, I will investigate frequent words in the tweets. Secondly, the links that were circulated in the tweets with the keyword "vaccine". Then, I will analyze the sentiment of the tweets that contain the biontech and sinovac as keywords. Finally, I peruse the network structure of the three information ecosystems of tweets with keywords vaccine, biontech and sinovac separately.

### 4.3.1 Frequency Analysis

Table 2, Table 3, and Table 4 show the top 10 words of the tweets with the "vaccine", "biontech", "sinovac" keywords, respectively. Additionally, the tables incorporate the English translation or the details of the words and the frequency and how many times that particular word appears in the tweets. Intuitively, the word vaccine is the most frequent word in the tweets of vaccine, biontech and sinovac keywords. Also, the words "china" and "turkey" are among the most frequent words that appear in each keyword. For the vaccine tweets, we encounter "Dr Fahrettin Koca" (health minister of Turkey Republic), "million doses" of vaccine orders, science, mask as the frequent words in Table 2. For the tweets with biontech keywords, we see the co-producer company Phizer the CEO Uğur Şahin and co-founder Özlem Türeci among the most frequent words in Table 3. Importantly, the frequent use of the words "German" and "Sinovac" gives us the insights that Turkish Twitter users care about the origin of the BioNTech vaccine and compare it to the Sinovac vaccine. When it comes to Sinovac tweets in Table 4, it distinguishes with the words "phase" and "brazil" basically, individuals frequently refer to the phase 3 result of the Sinovac vaccine in Brazil.

| Word | Details or translation | Frequency |
|---|---|---|
| aşı | vaccine | 337,977 |
| sağlık | health | 319,583 |
| drfahrettinkoca | health minister's name | 217,570 |
| milyon | million | 208,696 |
| doz | dose | 193,347 |
| çin | china | 163,659 |
| ilk | first | 153,411 |
| türkiye | turkey | 150,923 |
| bilim | science | 117,675 |
| maske | mask | 116,209 |

Table 2: Top-10 Words in Vaccine Tweets

| Word | Details or translation | Frequency |
|---|---|---|
| aşı | vaccine | 168,676 |
| phizer | phizer | 61,562 |
| uğur şahin | CEO of BioNTech | 31,378 |
| çin | china | 23,767 |
| ilk | first | 21,981 |
| türkiye | turkey | 21,849 |
| pfizer/biontech | pfizer/biontech | 19,576 |
| alman | german | 19804 |
| sinovac | sinovac | 19043 |
| özlem türeci | co-founder of BioNTech | 17783 |

Table 3: Top-10 Words in BionTech Tweets

| Word | Details or translatio | Frequency |
|---|---|---|
| aşı | vaccine | 109,799 |
| çin | china | 35,685 |
| türkiye | turkey | 21,974 |
| faz | phase | 21,648 |
| türkiye | turkey | 21,849 |
| sağlık | health | 18,332 |
| brezilya | brasil | 13,920 |
| drfahrettinkoca | health minister's name | 13,498 |
| çinli | chinese | 12881 |
| biontech | biontech | 11893 |

Table 4: Top-10 Words in Sinovac Tweets

#### 4.3.2 Shared Link Analysis

To facilitate understanding the vaccine information ecosystem on Twitter, I analyze the links shared in vaccine tweets. First, 59373 unique links were shared 106292 times during the period of the study. Table 2 shows the top ten domain addresses in the tweets containing the keyword "vaccine" (Turkish for "vaccine"). The most frequently shared domain is YouTube, with 4613 shares. The 8 out of 10 popular domains are news media sites ranging from independent news sources (birgun.net) to private media conglomerate (ntv.com.tr), from state-run news agency (aa.com.tr) international news sources (sputniknews.com). Future research might establish credibility scores to those domain addresses to assess how fast and deep the fake and true information spread across the information ecosystem [14].

| Domain | Frequency |
|---|---|
| youtube.com | 4613 |
| birgun.net | 3416 |
| sozcu.com.tr | 3280 |
| cumhuriyet.com.tr | 2435 |
| t24.com.tr | 1510 |
| pscp.tv | 1359 |
| aa.com.tr | 1267 |
| tele1.com.tr | 1253 |
| sputniknews.com | 1211 |
| ntv.com.tr | 1208 |

Table 5: The most frequent 10 domain addresses in vaccine tweets

### 4.3.3 Sentiment Analysis

I use the berturk social language model [23] to analyze the sentiment of the tweets. Figure 10 shows the weekly sentiment of biontech and sinovac tweets. Since the model's accuracy for positive tweets is not adequate (see Appendix A) I would conclude that the model can be improved by fine-tuning the model for the positive tweets on the vaccine tweets.

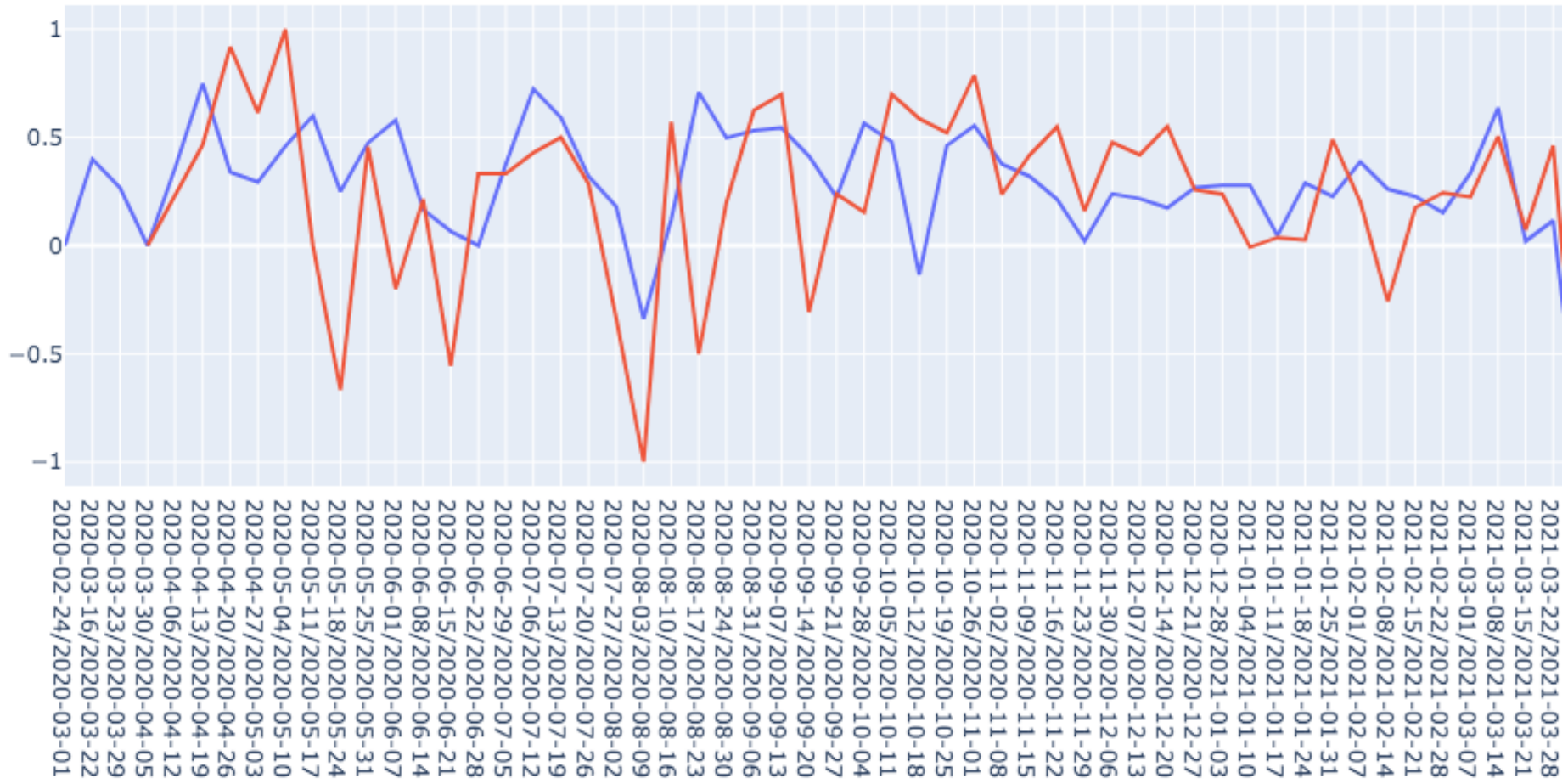

Figure 10: Sentiment of Biontech and Sinovac Tweets

## 4.4 Network Analysis of Vaccine Tweets

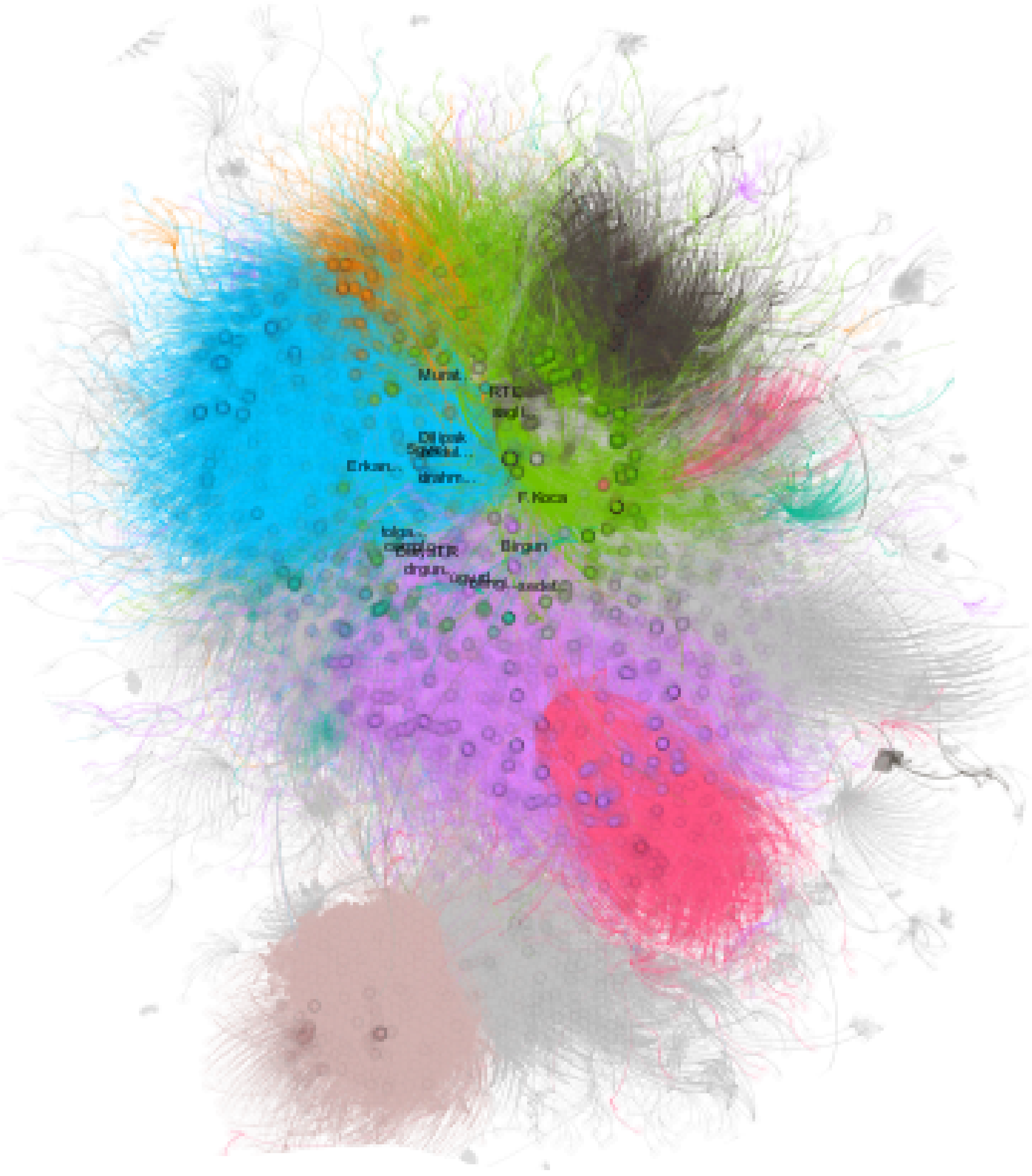

Figure 11: Network Graph of Vaccine Tweets

Figure 11 shows the network map of the accounts that are in the vaccine keyword information ecosystem. By using the Gephi Graph Visualization tool, I create the directed network map. Each node is the Twitter account, and each edge is the summation of the interaction of quotation, mention and retweet be-tween nodes. I colour the nodes by modularity class to detect communities, and node size is proportional to accounts eigenvector centrality.

Four communities were detect. The first one is green nodes of the health officials, incumbent party-affiliated media and the accounts interacting with them. Dr Fahrettin Koca's health minister is the central point of the green community's vaccine information ecosystem. This community constitutes 31% of the total nodes. On the other hand, the dark grey community, corresponding 6.5% of the nodes, represent the ruling party AKP's supporters. Recep Tayyip Erdoğan (@RTE) is the central node of this community. Thirdly, the purple nodes are medical doctors supporting the vaccine and the users that interact with them. They represent the 25% of the active users in the vaccine information ecosystem at Turkish Twittersphere during the study period. On the contrary, the blue nodes are the anti-vaxxers and conspiracy theorists, with 13% community share.

To understand the topic of conversation that took place in each of the four communities, LDA [28] topic modelling is implemented to central nodes in the four communities. The eigenvector centrality score detects central nodes. Then, tweets of the 12 nodes for the conspiracy theorist nodes and 11 nodes for the rest of the three communities are used for topic modelling.

### 4.4.1 Topic Modeling

Table 6 illustrates the topic modelling of the conversation that took place in different communities. The supporters of the ruling party focus on the opportunity for the political campaign during the pandemic. For instance, they emphasize million-dose deals of vaccine with Sinovac and BioNTech, vaccination services being free of charge to underline how Turkey Turkey is good at vaccination compared to "world".

| Group | Focus Topic | Words in LDA | Turkish |
|---|---|---|---|
| AKP supporters | COVID-19 Management Campaign | free vaccine, first in the world in vaccination | bedava, ücretsiz aşı, aşılama dünyada ilk |
| Health officials | Briefing | COVID-19 cases, number of vaccines ordered | milyon doz, yaş, canlı |
| Medical doctors | Explaining the Science | phase study, first dose of vaccine | faz çalışması, ilk doz |
| Anti-vaxxers | Denouncing Science and Pandemic | fake science, bill gates | olmayan, sahte bilim |

Table 6: LDA Topic Modelling of 4 Communities

On the other hand, health officials mainly make a press release by announcing the vaccine orders from Sinovac and BioNTech, and the vaccination schedule regarding age and chronic disease. Thirdly, medical doctors explain the phase study results, how effective the vaccines are against coronavirus. Last but not least, anti-vaxxers undermine the existence of the coronavirus pandemic. They advocate that coronavirus is not real. They believe that the virus was either produced in the lab environment by the globalist scientist and/or the coronavirus pandemic is just as dangerous as ordinary influenza rather more dangerous. Therefore, they tend to use the term "fake pandemic". Furthermore, they believe that globalist actors such as Bill Gates aim to change the creation of humankind via mandatory vaccination. Consequently, they reject the pandemic preventive measures such as the face mask requirement and mandatory vaccination. Those conspiracy theorists denigrate the statements of the scientific community by the term "fake science".

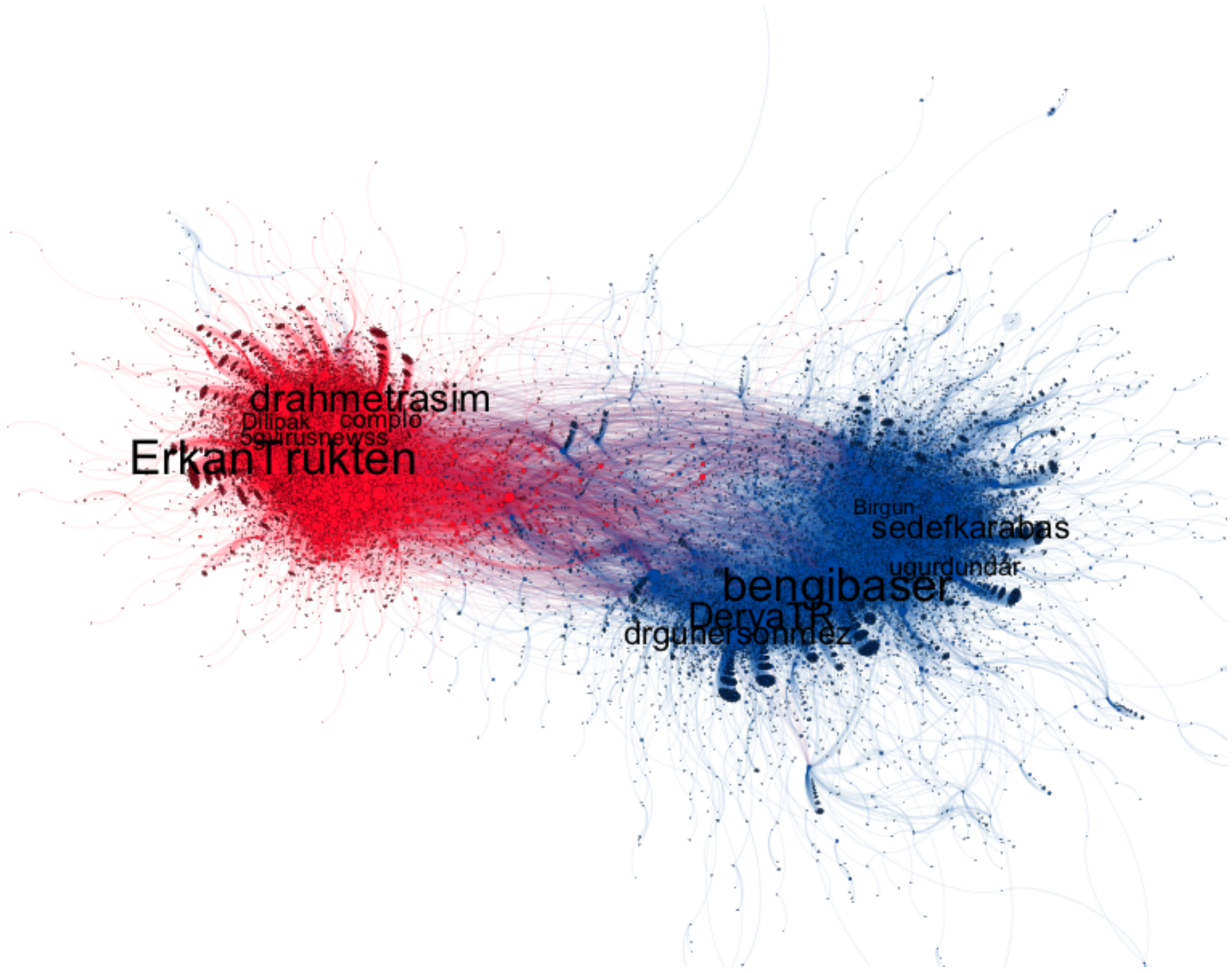

Figure 12: Anti-Vaxxers vs. Vaccine Supporters

Figure 12 illustrates the network visualization of the anti-vaxxers and medical doctors. The red community is the anti-vaxxers, and the blue one is the medical doctors. The names are the central nodes of these communities. It is important to notice how polarized these two communities are, suggesting the users belong anti-vaxxers community does not encounter the tweets of the medical doctors.

# 5 Discussion & Future Research

In the first part of this study, I investigate the public opinion change in several important themes on Turkish Twittersphere using quantitative text analysis methods. It provides a comprehensive research design to analyze the public opinion change in important events such as a pandemic.

Intuitively, the COVID-19 pandemic has occupied the public discussion with the problem it brought. However, it is important to highlight the pre-COVID-19 public discussion, the change in the number of tweets as a reaction to governmental actions and how rapidly public opinion could change due to critical events such as a pandemic. Importantly, social bots are detected to be prevalent in COVID-19 discussion on Turkish Twittersphere.

In the second part, this study analyzes the information ecosystem regarding the vaccine on the Turkish Twittersphere. First, it shows the change in the daily volume of the vaccine tweets. Furthermore, the social network analysis tools detect important communities existing in the vaccine ecosystem. Having investigated the content of the different communities' tweets shows the diverging agenda of the groups. It has vital importance since vaccine is one of the most controversial topics of the pandemic. While some regard it as an important cure, others are hesitant to be vaccinated and anti-vaxxers, deliberately reject it. Therefore, understanding the rationale behind (mis)trust in the vaccine would help researchers, government officials to find an efficient way to convey their message to the public.

Although the role of alternative information sources like social media is highly increasing in Turkey, the most prominent news source is still traditional media outlets [29]. Especially, the rate of using television as the sole source of news is much higher in the elderly. This makes them dependent on the accuracy of TV news since they do not use any fact-checking systems or social media as an alternative news course. Therefore, studies should also consider the offline information ecosystem to understand the source of mis/disinformation.

It should also be noted that offline traditional media is also problematic as an information source. For instance, rumours were circulating in TV channels that Turkish genes can resist the virus before the outbreak of the COVID-19 in Turkey. And, some TV channels were announcing that traditional Turkish foods protect against the virus in their prime times. Those rumours are not the words of the floor show, but a medical practitioner discusses them on national TV channels.

Future research would advance Turkish natural language processing. This research does not use lemmatisation and stemming since there is no adequate programming library for Turkish language. Also, developing an efficient sentiment model for vaccine tweets would be a potential indicator to control the vaccine's mis(trust) level. Last but not least, distinguishing the credible information sources from the uncredible ones would serve to monitor the potential mis/disinformation circulating on social media platforms .

# A Hand Coding for Sentiment

| Tweet | Hand-Coded Sentiment |
|---|---|
| Hocam Sinovac bile genel koruyuculuğu %50 diyordu. Nasıl %80 çıkıyor | NEG |
| @saglikbakanligi @drfahrettinkoca yahu daha 3 ncü faz çalışması yapılmamış Çin malı sinovac aşısı için nasıl %91,25 etkili açıklama yapabiliyorsunuz. Biontech aşısı neden Türk milletine layık görmüyorsunuz? | NEG |
| @DrGunerSonmez sinovac aşısı yaptırdıktan sonra biontech aşısı yaptırilabilinirmi .yaptırılabilir ise ne kadar zaman geçmesi lazım? | NOTR |
| Ya ben de çok güvenmiyorum aslında sinovac'a. Ama sıramız gelirse oluruz mecbur | NEG |
| @HaberturkTV Devlet sinovac aşısı güvenli, bunu olabilirsiniz dedikten sonra bizim devlete güvenmeme gibi bir durumumuz söz konusu olamaz, olmamalı. Devlet bizim asla kötülüğümüzü istemez. Devletimize güvenimiz tam! Cin aşısını gönül rahatlığıyla olacağız inşallah. | POS |
| KobayDeğil AşıOlalım Hiç değilse Rus aşısı alalım. En azından kendi insanı için hali hazırda kullanıyor. Sinovac olmaz! | NEG |
| Türkiye'nin de aldığı Çinli #Sinovac aşısının Brezilya'da yüzde 78 etkili olduğu açıklandı | NOTR |
| @timbooth75 Çin sinovaç RNA aşısı değilmiş .tercih edilmelimi ? | NEG |
| @ugurdundarsozcu aaa grip aşısı da mı Sinovac mış? | NEG |
| @bengibaser Bizim ki sinovac değil miydi | NOTR |

Table 7: Tweets classified as Negative Sentiment by BerTurk-Social Model

| Tweet | Hand-Coded Sentiment |
|---|---|
| Pfizer: Geliştirdiğimiz aşı eksi 70 derecede saklanmalı Moderna: Geliştirdiğimiz aşı -20 derecede saklanmalı Çin SinoVac: Aşımız kısık ateşte de saklanabilir<br>O zaman en güvenilir Çin aşısı | NEG |
| Bugün gönüllü olarak Sinovac için 8 ocak tarihine randevu aldım bakalım hayırlısı. | NOTR |
| Çin'de Sinopharm firması tarafından üretilen aşı, resmî onay alan ilk aşı oldu. Türkiye'nin de aldığı Sinovac ise, nihai onay için kobay olarak kullandığı Türkiye, Brezilya ve Endonezya'daki üçüncü faz deneme sonuçlarını bekliyor.. | NOTR |
| Reuters: Sinovac'ın aşısı ileri aşama denemelere göre yüzde 78 etkili | NOTR |
| Sn. Karadağ, Sinovac aşısının Türkiye ve Azerbaycan distribütör,ü olan şirketi araştırırsanız, bir yerlere varabilirsiniz, belki! | NOTR |
| Sağlık Bakanı'ndan Sinovac açıklaması:<br>"Etkinlik yüzde 83,5 ve hastaneye yatışları yüzde 100 engelliyor" | NOTR |
| En azından Avustralya'da alternatif var. TR de Sinovac işine gelirse | NEG |
| Koca: Başından beri bu üç aşı için faz-3 çalışmasını başlatmak istedik. Biontech başladı, Sinovac'ın faz-3 çalışması başladı, hatta biz destek verdik | NOTR |
| Harvard Üniversitesi'nden Doç. Altındiş: Risk altında olsam, tek seçeneğim Sinovac aşısı olsa bu aşıyı olurdum. | NOTR |
| Son iki ekran görüntüsünün linkini bırakıyorum. Diğer aşı ve ilaç denemesi yapılmış isimlerine bakın belki kullandıklarınız vardır. | NOTR |

Table 8: Tweets classified as NOTR Sentiment by BerTurk-Social Model

| Tweet | Hand-Coded Sentiment |
|---|---|
| Bu zibidiyi buraya çıkardığıniz için bu gece sizi protesto ediyor ve izlemiyorum... | NEG |
| @Gokhan9oglu @drfahrettinkoca Siyasi! Kendi birliğinin ürettiği aşı ve ABD baskısı ;) Sinovac geleneksel (grip aşısı) yöntemlerle üretilmiş,daha güvenilir bir aşı | POS |
| @BulutGulcuN sinovac aşısının güvenilirliği ile ilgili bilgilendirme. may be | NOTR |
| Hazır aşılar gündemdeyken.. aşı Sinovac çinaşısı zam | NEG |
| %50 etkili sinovac aşısı bünyede covid19la denk gelirse https://t.co/zhAvYkP8Tk | NOTR |
| Çin, Sinovac yerine önce Sinopharm'ın Covid-19 aşısına izin verdi | NOTR |
| Sinovac (Çin aşısı) | NOTR |
| Bakan @drfahrettinkoca, sizin bu Sinovac ile gizli bir ortaklığınız mı var? Sahtekarlığı kanıtlanmış bir firmanın güvenilirliği olmayan covid aşısında neden bu kadar ısrarcısınız? Unutma Sn. Bakan, yalancının mumu yatsıya kadar yanar... | Neg |
| Bilim Kurulu, Çin malı Sinovac aşısının etkinliğini yüzde 91,25 olarak açıklayalı 14 gün,<br>Brezilya'daki faz-3 çalışmalarında bu aşının koruyuculuğu yüzde 78 olarak tespit edileli birkaç saat oldu. | NEG |
| Türkiye, KKTC'ye 20 bin doz Sinovac aşısı gönderdi | NEG |

Table 9: Tweets classified as Positive Sentiment by BerTurk-Social Model

## B Shared Links

| Domain | Frequency |
|---|---|
| youtube.com | 4613 |
| birgun.net | 3416 |
| sozcu.com.tr | 3280 |
| cumhuriyet.com.tr | 2435 |
| t24.com.tr | 1510 |
| pscp.tv | 1359 |
| aa.com.tr | 1267 |
| tele1.com.tr | 1253 |
| sputniknews.com | 1211 |
| ntv.com.tr | 1208 |
| instagram.com | 1082 |
| bbc.com | 1075 |
| sabah.com.tr | 1047 |
| yenicaggazetesi.com.tr | 1029 |
| gazeteduvar.com.tr | 1026 |
| sol.org.tr | 981 |
| indyturk.com | 977 |
| sondakikaturk.com.tr | 964 |
| halktv.com.tr | 905 |
| odatv4.com | 793 |
| habervakti.com | 783 |
| bizimtv.com.tr | 773 |
| gercekgundem.com | 734 |
| tevhidgundemi.com | 721 |

Table 10: The most frequent 30 domain addresses in vaccine tweets

# C Word Clouds

Covid Once:

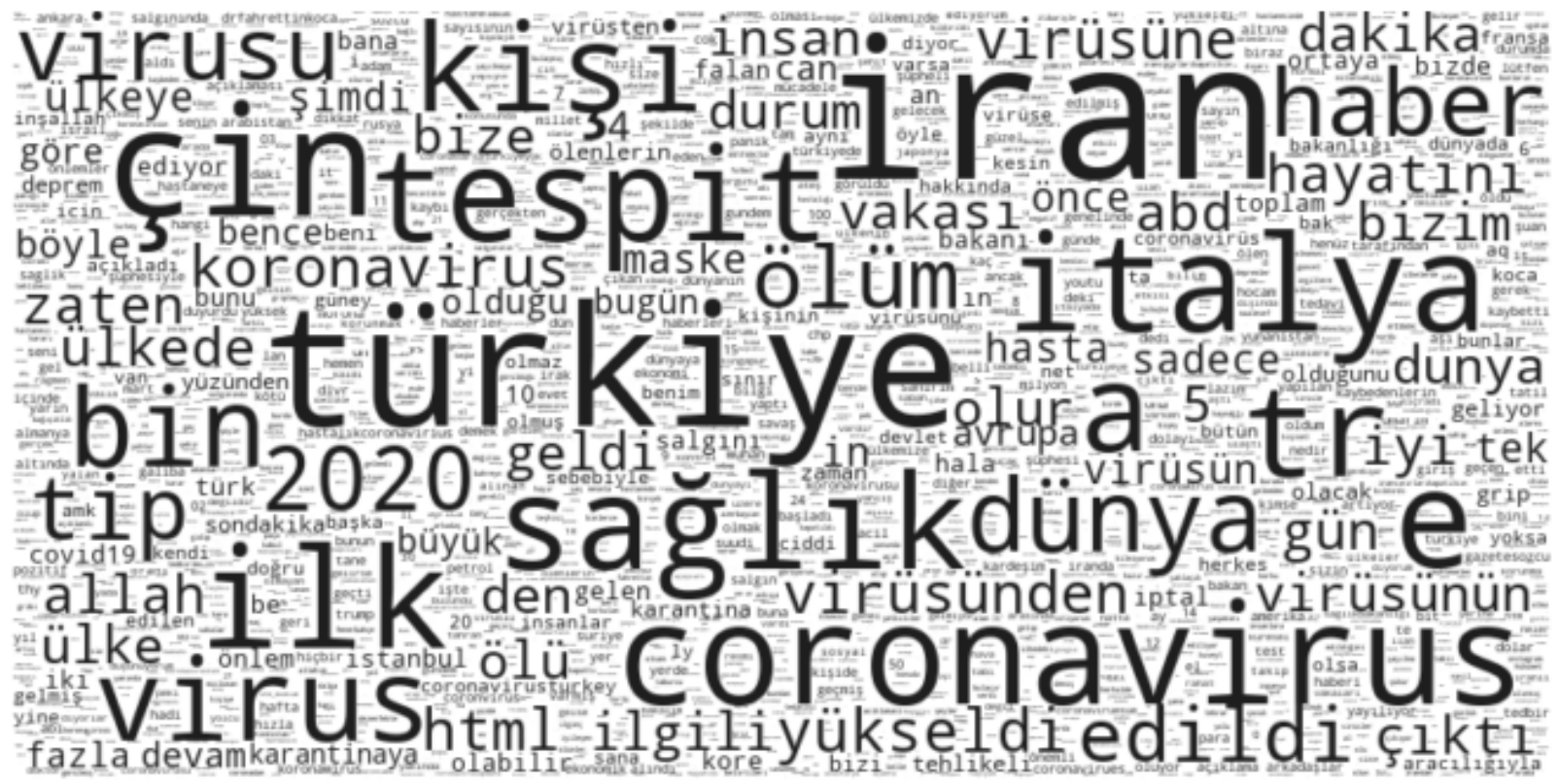

Figure 13: COVID-19 Before

Covid Sonra:

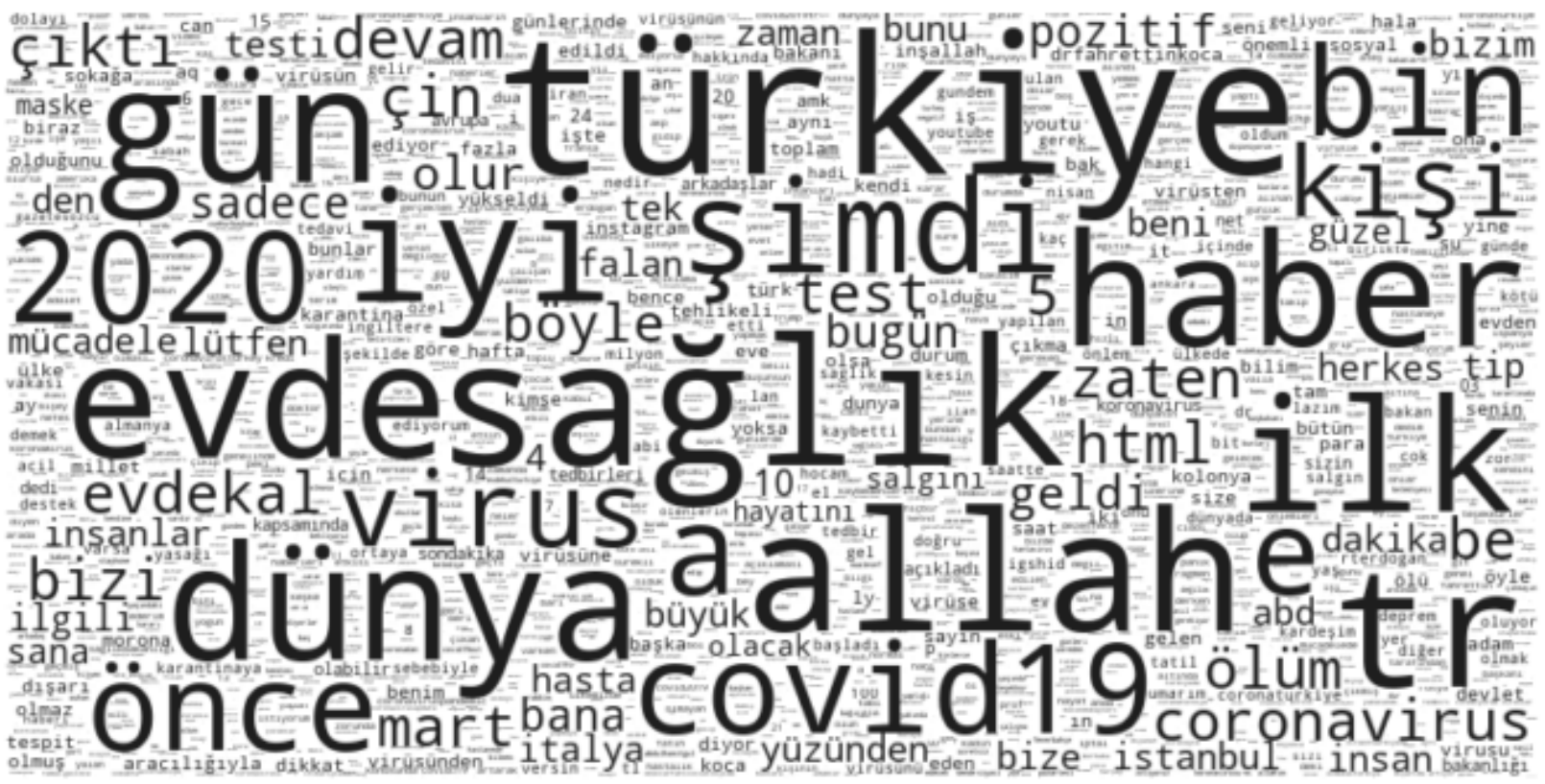

Figure 14: COVID-19 After

Politika Once:

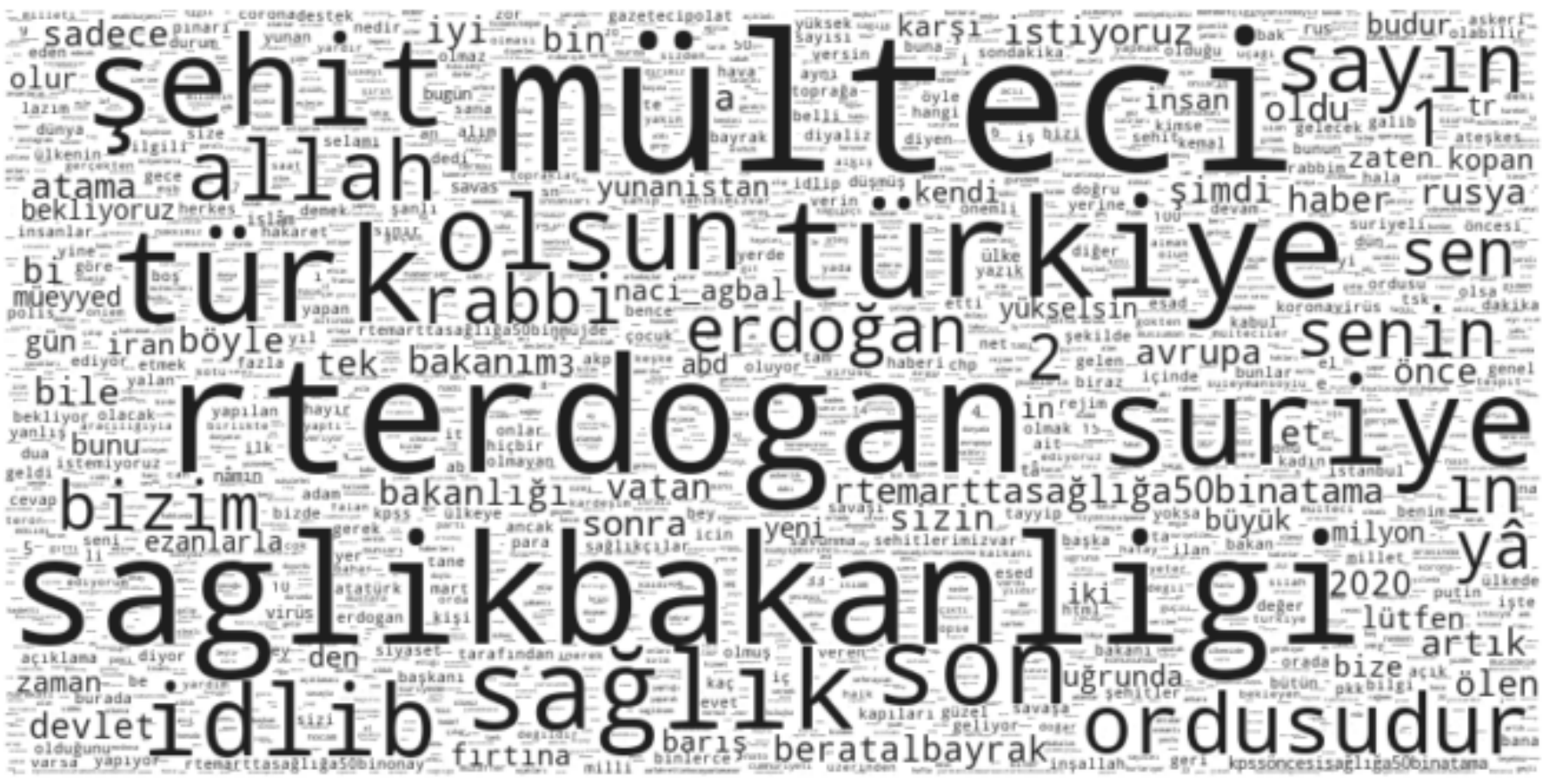

Figure 15: Politics Before

Politika Sonra:

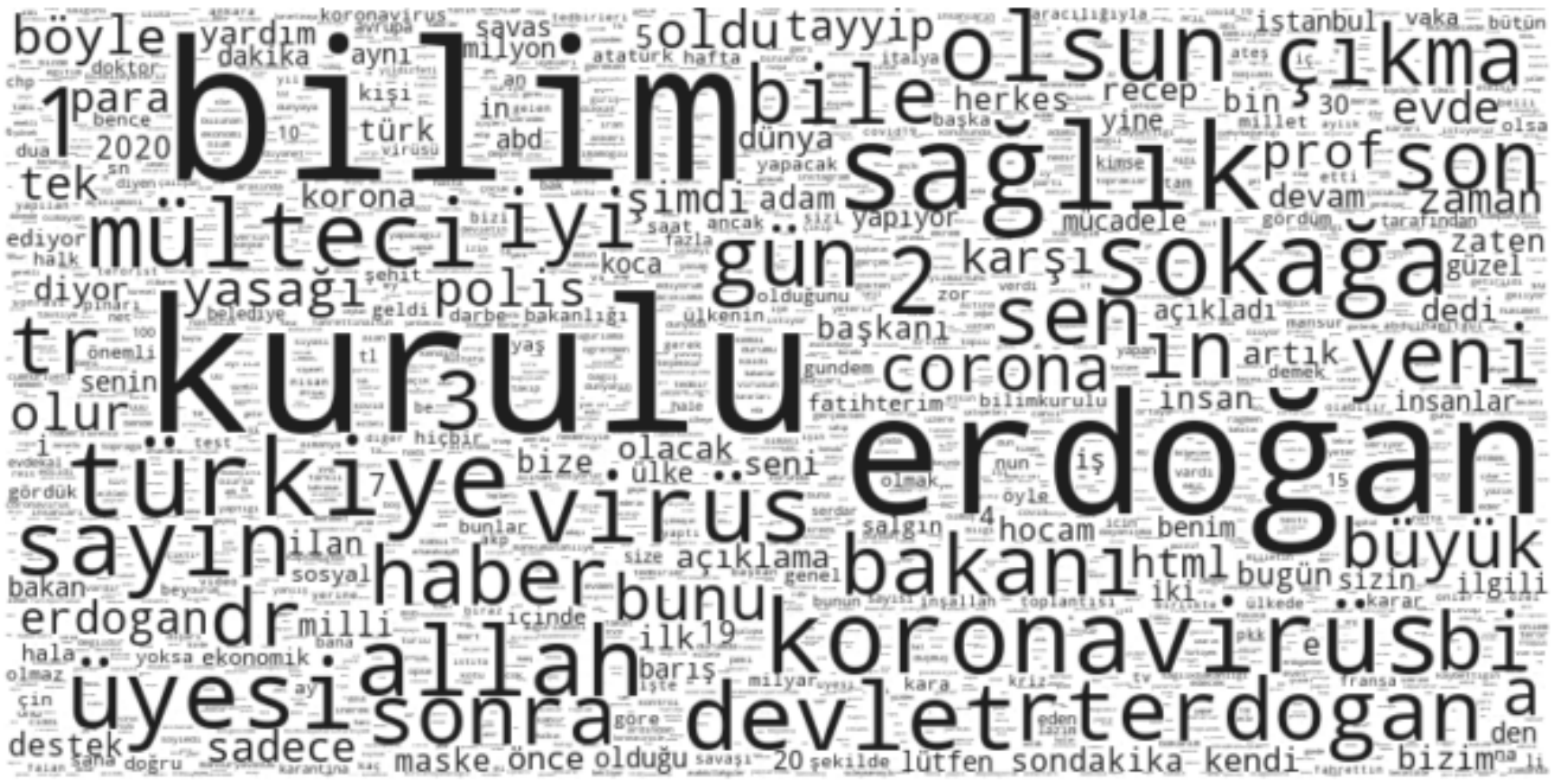

Figure 16: Politics After

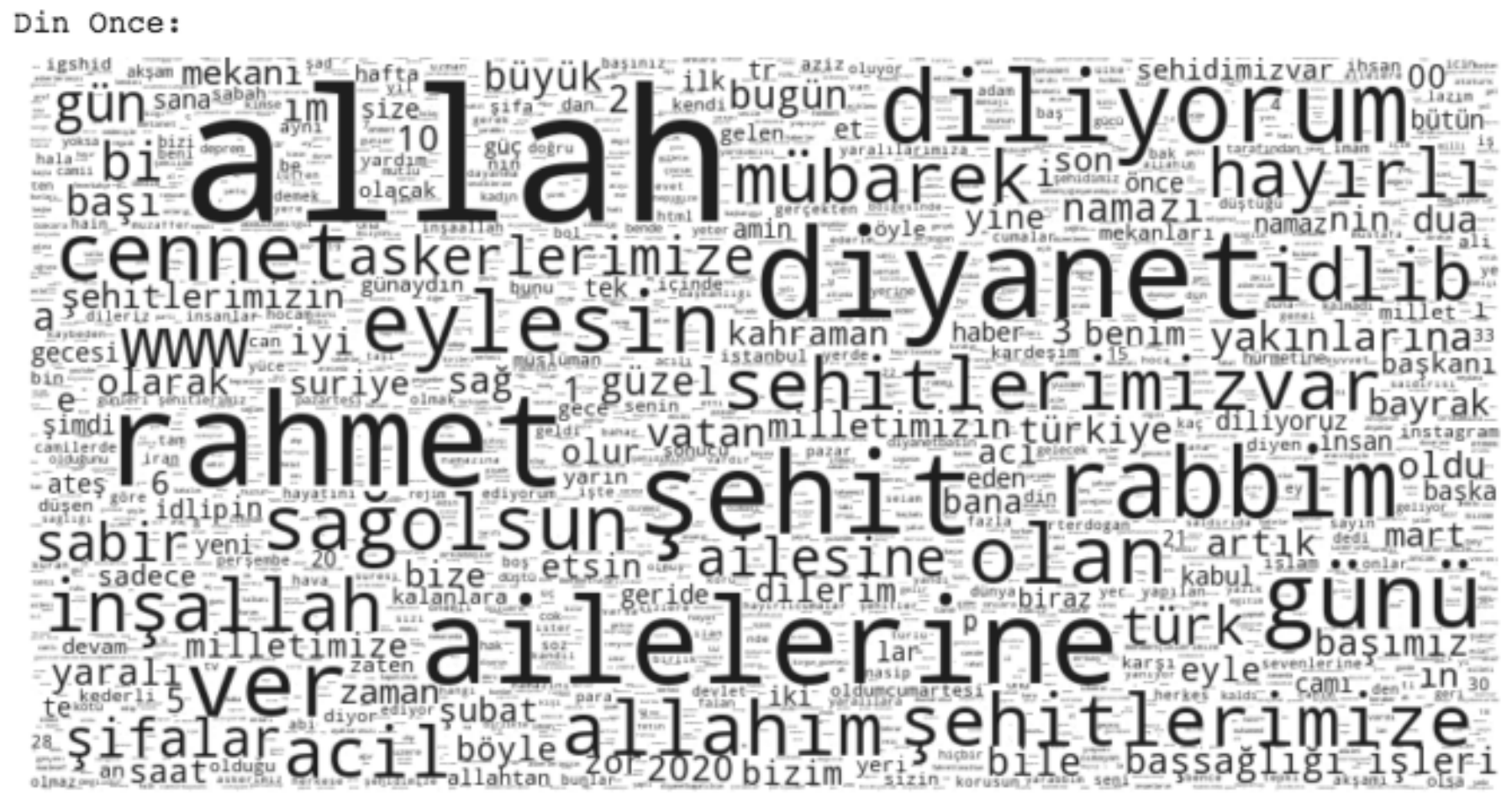

Figure 17: Religion Before

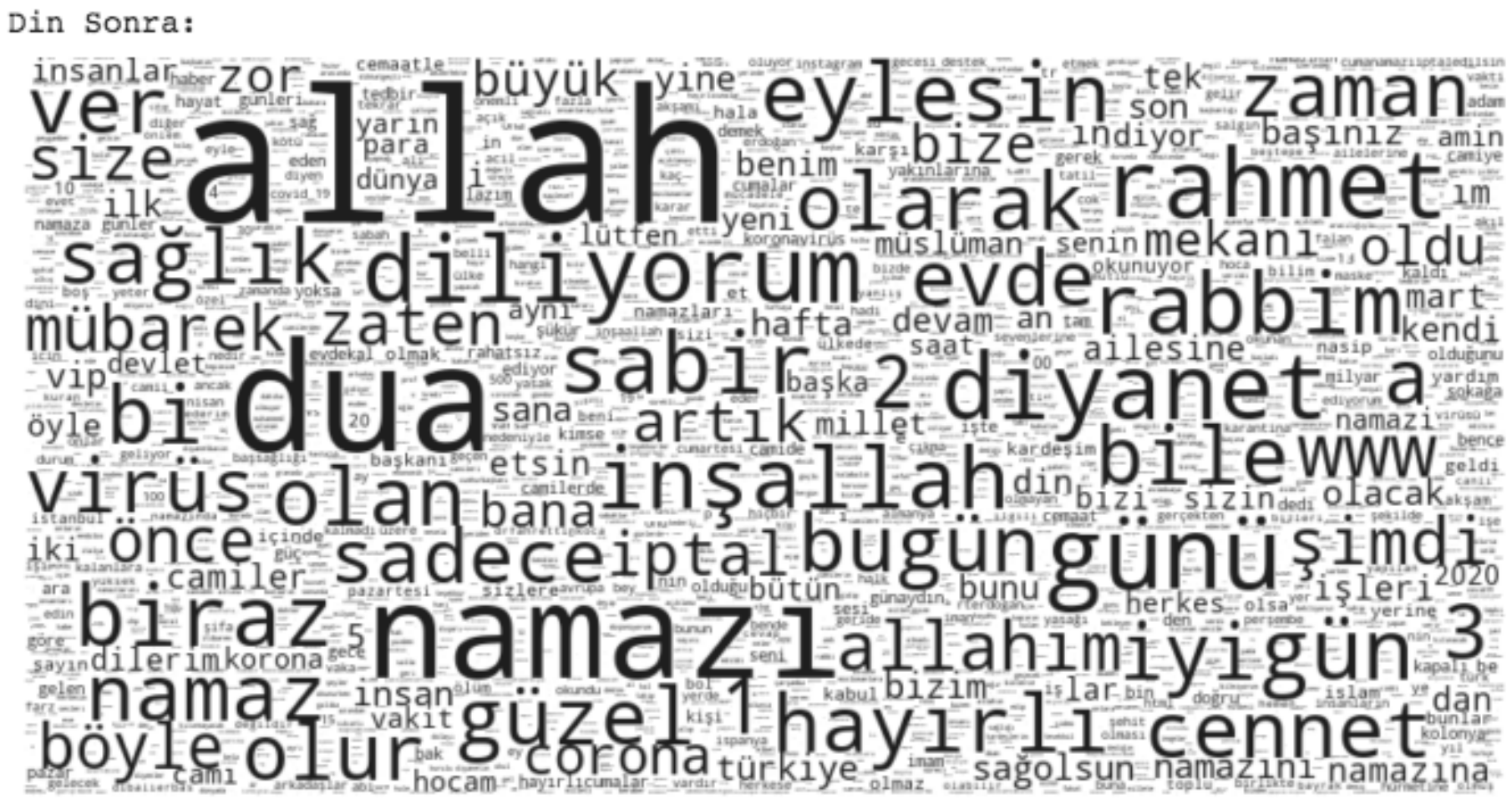

Figure 18: Religion After

Ekonomi Once:

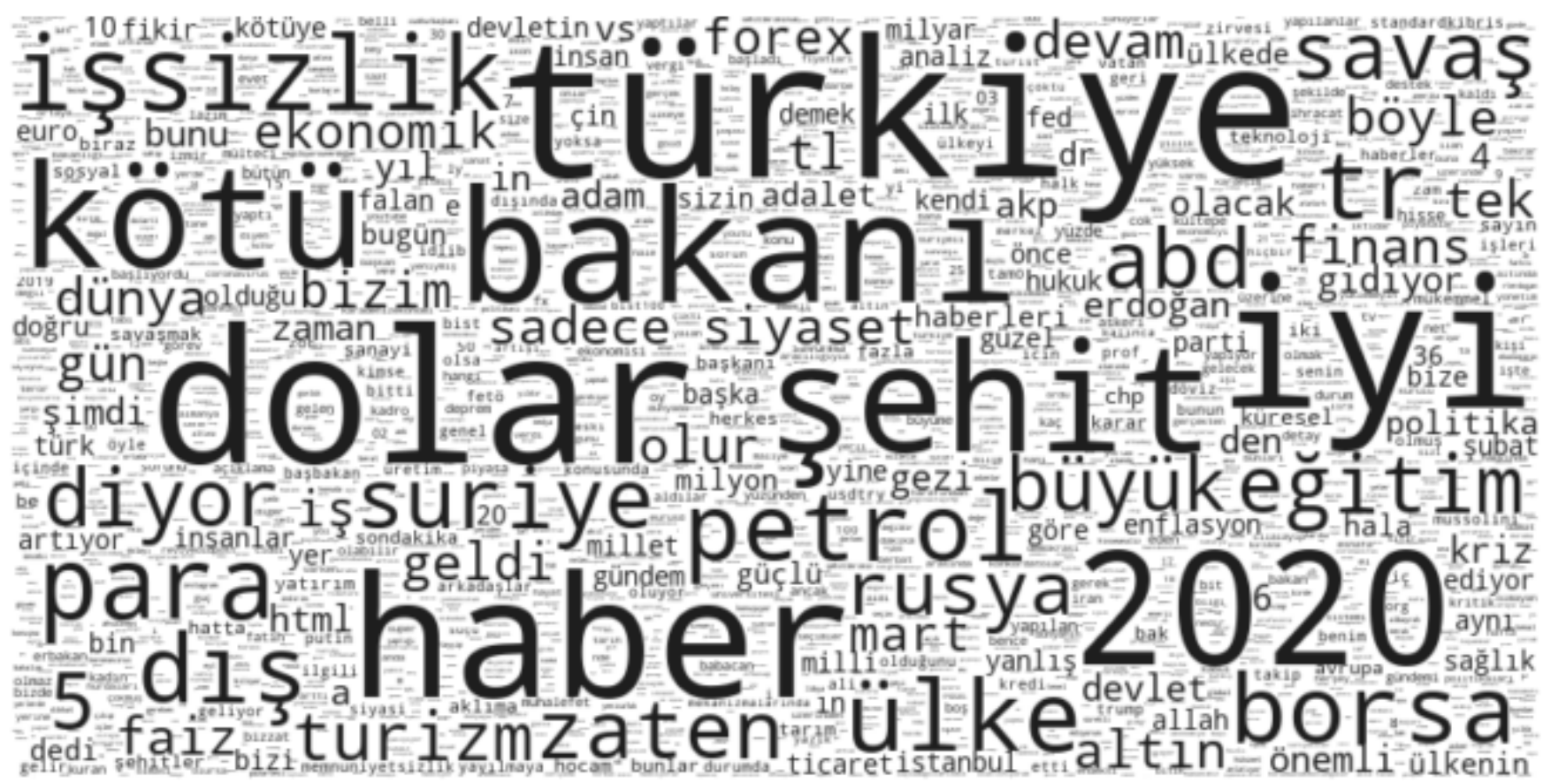

Figure 19: Economy Before

Ekonomi Sonra:

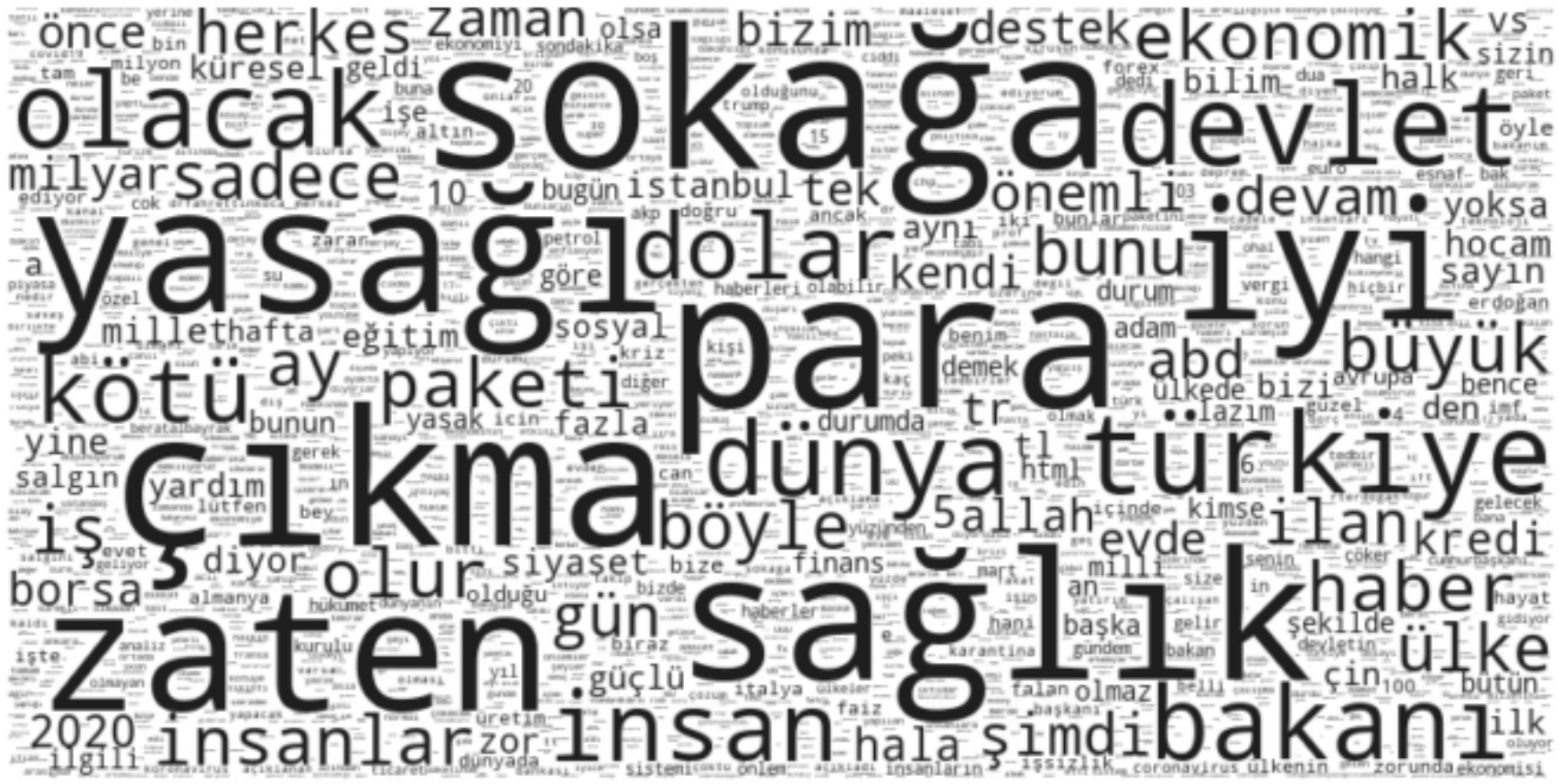

Figure 20: Economy After

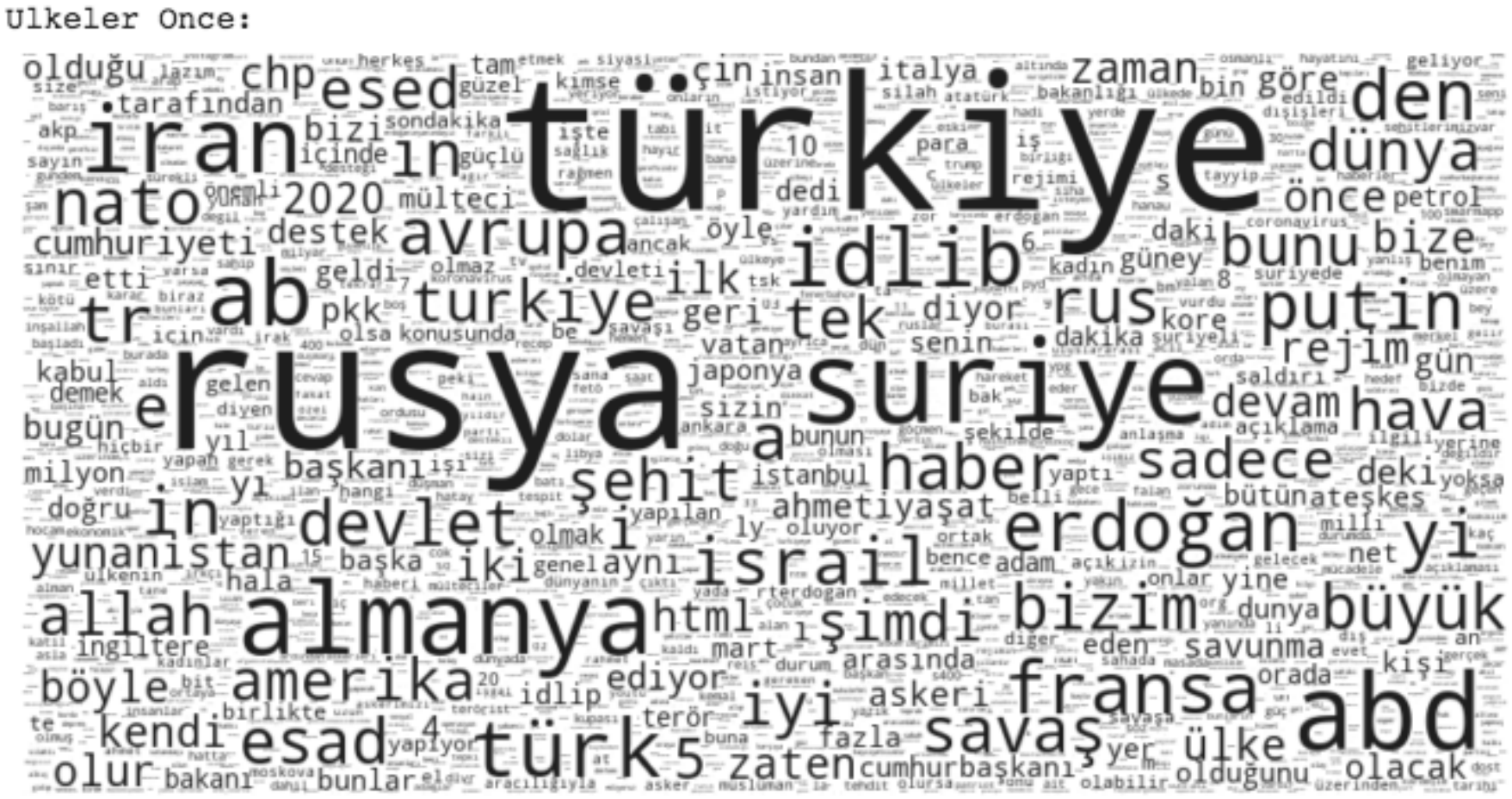

Figure 21: International Relations Before

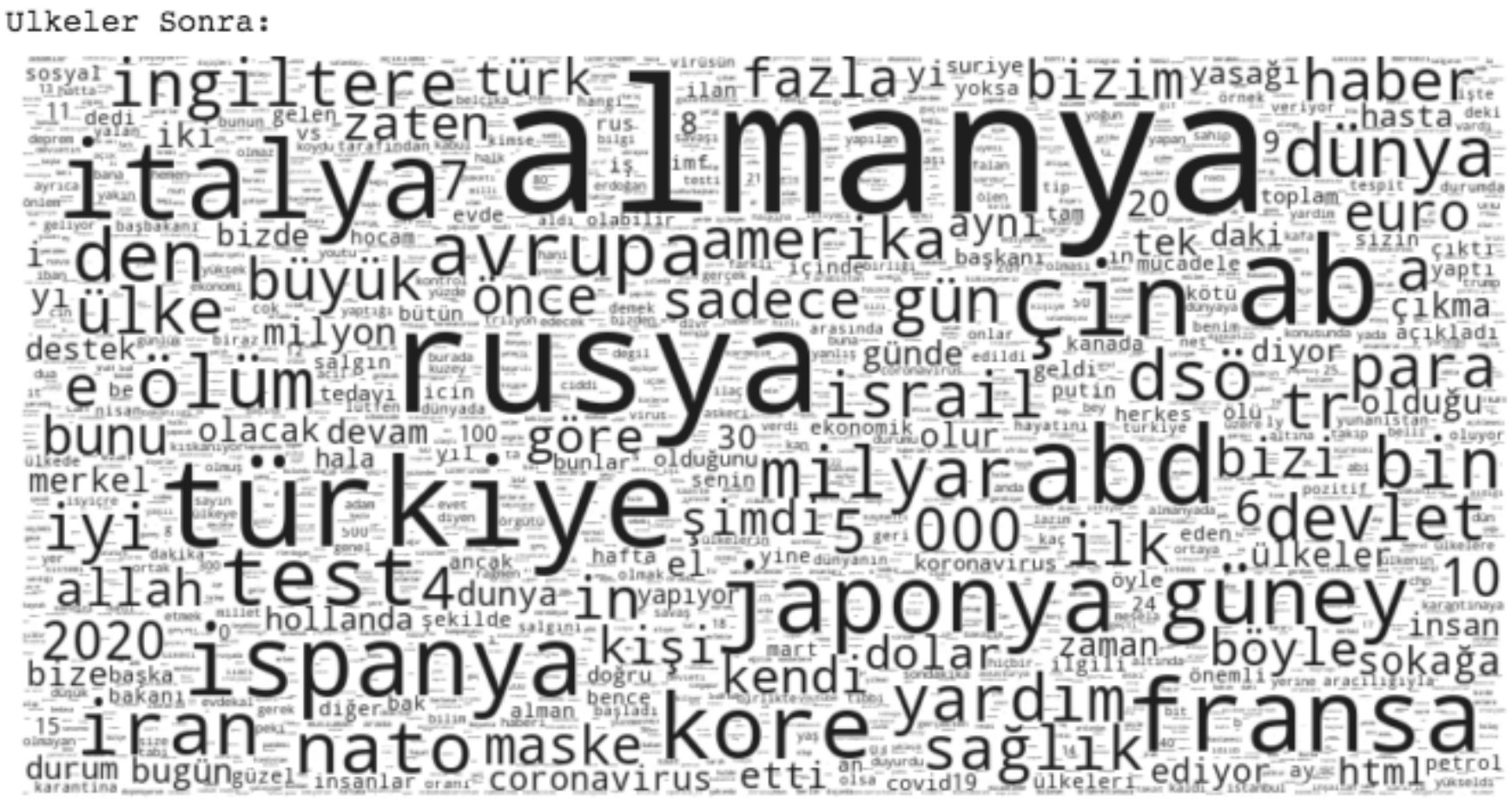

Figure 22: International Relations After

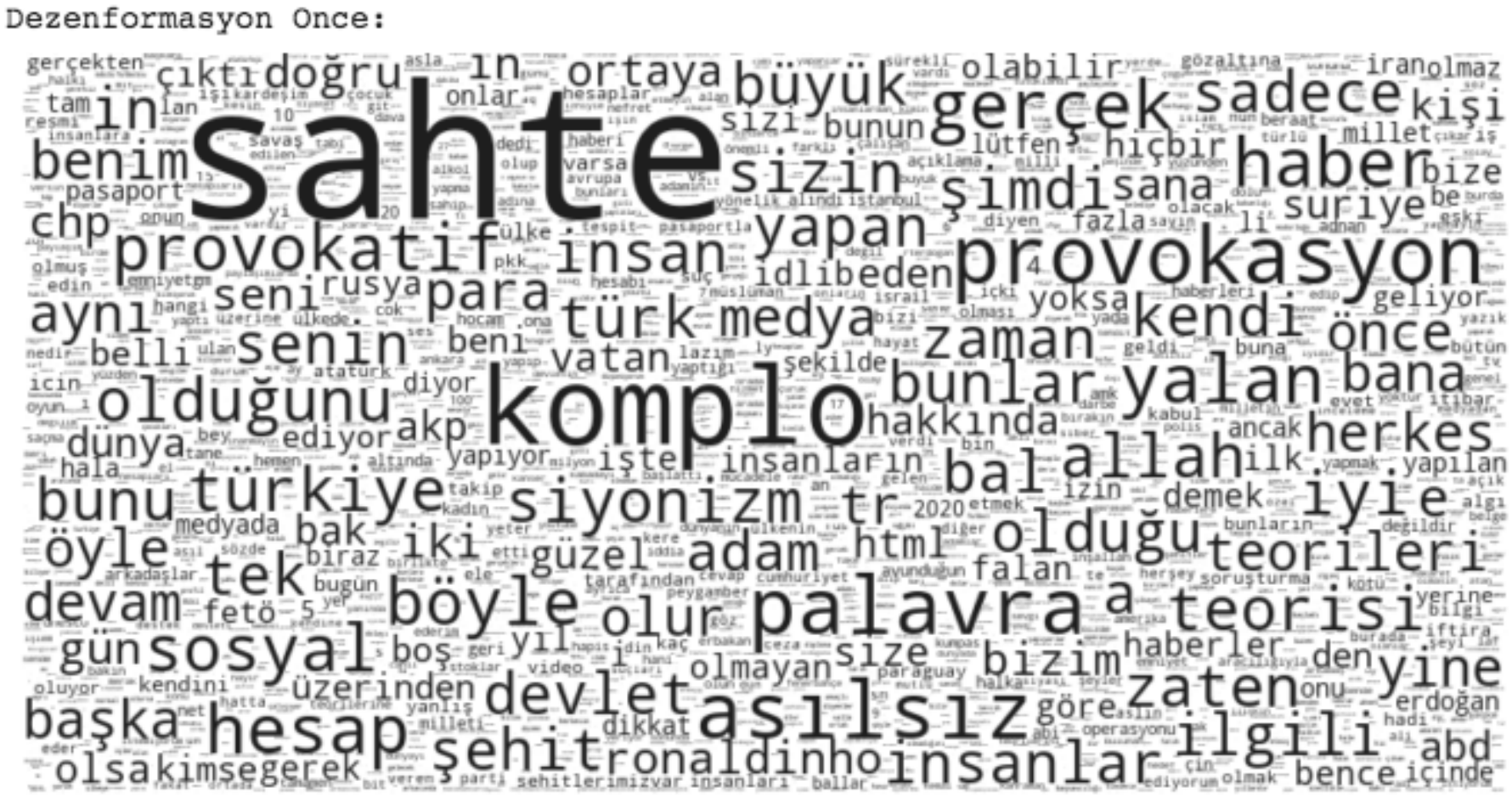

Figure 23: Disinformation Before

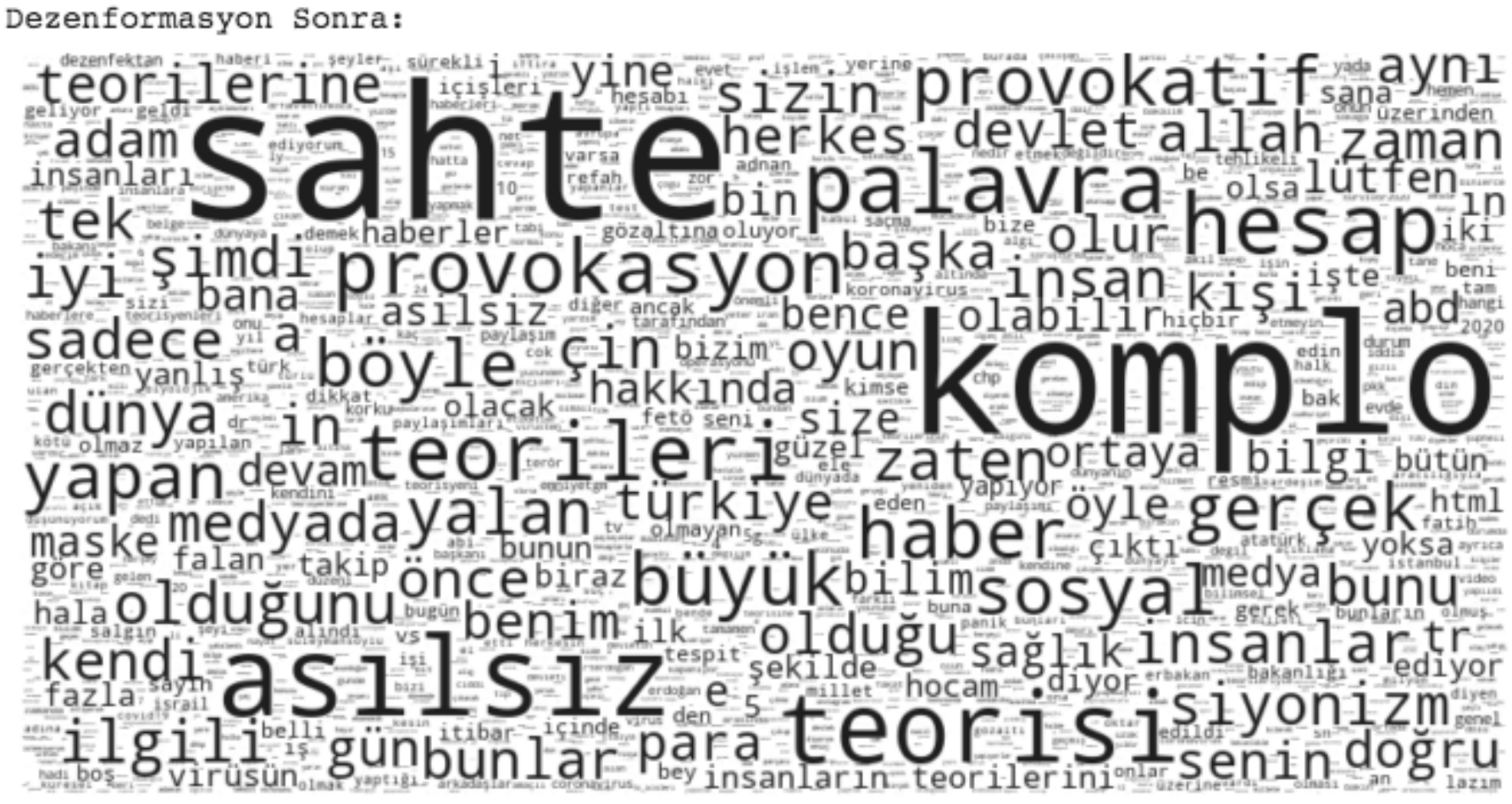

Figure 24: Disinformation After

# D Network Visualizations

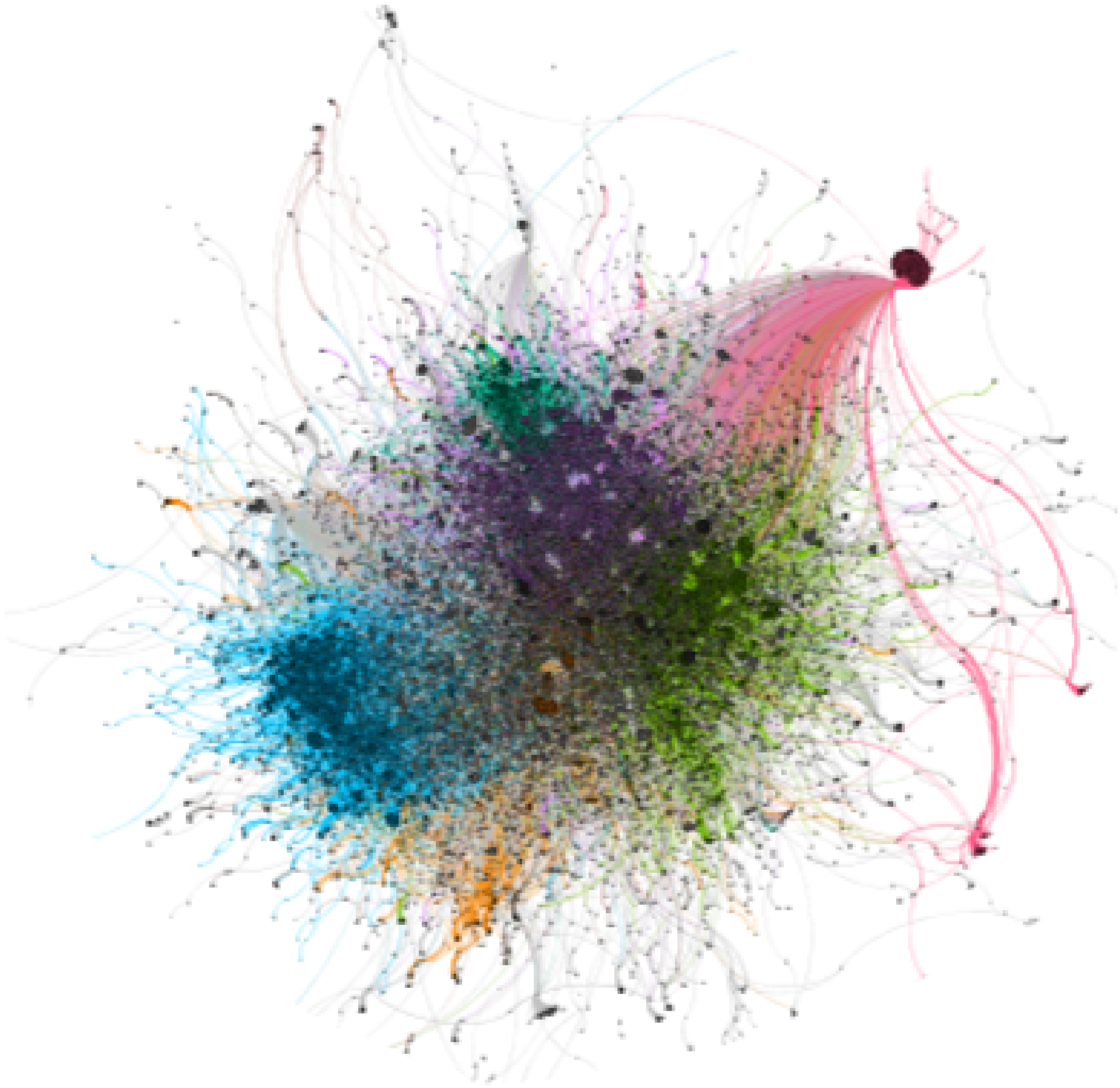

Figure 25: Network Graph of Biontech Tweets